\pdfoutput=1

\documentclass[10pt,aps,prb,final,twocolumn,amsmath,amssymb,groupedaddress,longbibliography]{revtex4-2}

\usepackage[pdftex,colorlinks=true,linkcolor=blue,citecolor=blue,urlcolor=blue]{hyperref}
\usepackage{dutchcal} 
\usepackage{aurical}
\usepackage[T1]{fontenc}
\usepackage{units}
\usepackage{braket}
\usepackage{graphicx}
    \graphicspath{{./../Figures/}}
\usepackage{amsmath,amssymb}
\usepackage{amsfonts}
\usepackage{mathrsfs}
\usepackage{epsfig}
\usepackage{dcolumn}
\usepackage{enumerate}
\usepackage{amsthm}
\usepackage{bbold}
\usepackage{subfigure}
\usepackage{tikz}

\usepackage{color,soul}
\usepackage{dutchcal}

\begin{document}

\title{Semiclassical propagation of coherent states and wave packets: hidden saddles}

\author{Huichao Wang and Steven Tomsovic}
\affiliation{Department of Physics and Astronomy, Washington State University, Pullman, WA. USA 99164-2814}

\date{\today}

\begin{abstract}
Semiclassical methods are extremely important in the subjects of wave packet and coherent state dynamics.  Unfortunately, these essentially saddle point approximations are considered nearly impossible to carry out in detail for systems with multiple degrees of freedom due to the difficulties of solving the resulting two-point boundary value problems.  However, recent developments have extended the applicability to a broader range of systems and circumstances.  The most important advances are first to generate a set of real reference trajectories using appropriately reduced dimensional spaces of initial conditions, and second to feed that set into a Newton-Raphson search scheme to locate the {\it exposed} complex saddle trajectories.  The arguments for this approach were based mostly on intuition and numerical verification.  In this paper, the methods are put on a firmer theoretical foundation and then extended to incorporate saddles {\it hidden} from Newton-Raphson searches initiated with real trajectories.  This hidden class of saddles is relevant to tunneling-type processes, but a hidden saddle can sometimes contribute just as much as or more than an exposed one.  The distinctions between hidden and exposed saddles clarifies the interpretation of what constitutes tunneling for wave packets and coherent states in the time domain.

\end{abstract}

\maketitle

\section{Introduction}
\label{intro}

The evolutions of Glauber coherent states for bosonic many-body systems and the mathematically nearly-identically related Gaussian wave packets arise in a multitude of physical contexts.  Examples abound in quantum optics~\cite{Glauber63,Scully97}, far out-of-equilibrium dynamics in bosonic many-body systems~\cite{Greiner02b,Polkovnikov11}, molecular spectroscopy~\cite{Heller81b, Gruebele92}, femto-chemistry~\cite{Zewail00}, and attosecond physics~\cite{Agostini04}.  Theoretical work encompasses, for example, coherent state representations of path integrals~\cite{Klauder85} and in the context of many bosons or the short wavelength limit, the semiclassical approximation~\cite{Klauder85,Huber88,Baranger01}.   This approximation is fundamental to studies of quantum-classical correspondence~\cite{Berry79b}, and pre- and post-Ehrenfest-time-scale dynamics~\cite{Heller75,Heller93,Tomsovic18} as well.

A complete semiclassical approximation for coherent state dynamics may be obtained by the saddle point approximation applied to coherent state path integrals~\cite{Klauder85,Baranger01}.  In the context of wave packets the essentially identical approximation is known as generalized Gaussian wave packet dynamics (GGWPD) and it has been proven to be equivalent to a complexified version of time-dependent WKB theory~\cite{Huber88}.  In either case, implementation in practice leads to considerable technical difficulties for any system possessing nonlinear dynamics.  

To begin with,  the classical trajectories that define the saddles' properties are solutions of a two-point boundary value problem, which is highly nontrivial for nonlinear dynamical systems.  If there are many degrees of freedom, this problem may be effectively impossible to solve.  It appears that for any propagation time $t\ne 0$, there exists an infinity of saddles; see Figure 1 ahead.  At the shortest evolution time scale, for minimum uncertainty wave packets (coherent states) only one saddle is physically relevant.  For a chaotic dynamical system, this time scale is extremely short and has a logarithmic dependence on $\hbar$.  The time scale is much longer for an integrable system, scaling algebraically in $\hbar$.   Nevertheless, in either situation this defines an Ehrenfest time regime~\cite{Ehrenfest27}.  Parenthetically, physically relevant means: i) the saddle is on the correct side of the Stokes lines; and ii) its contribution to the wave function at the relevant position is significant enough to be larger than any of the errors due to the saddle point approximations from other saddles contributing to the same position.  As time increases beyond the Ehrenfest time, the number of physically relevant saddles increases, necessarily creating quantum interferences, but the number is always finite at some fixed propagation time.  One might say that a set of measure zero saddles from the infinite set must be selected as the physically relevant saddles.  Furthermore, these trajectories necessarily involve complexified position and momenta.  This analytic continuation of real classical dynamics has many nontrivial features, such as branch cuts associated with singular runaway trajectories~\cite{Huber88}, and prefactor singularities that pose new problems for determining the phase (see the companion paper to this one ahead~\cite{Wang21b}).  

Considerable progress has been made in developing a practical method of identifying physically relevant saddles directly without encountering any of the irrelevant saddles~\cite{Vanvoorhis02,Vanvoorhis03,Pal16,Tomsovic18b}.  The basic idea builds on earlier techniques of identifying a single real reference trajectory for each classical transport pathway that exists in the nonlinear dynamics of the system~\cite{Oconnor92,Tomsovic93,Tomsovic93b,Barnes94}.  Then using a Newton-Raphson algorithm, a unique saddle point is identified for each classically allowed transport pathway and it accounts for that pathway's contributions to the dynamics.  The techniques have been developed using Gaussian wave packets, but it applies equally well to bosonic coherent states expressed in quadratures; see~\cite{Tomsovic18}.  In many cases the method can be extended to systems with many degrees of freedom by identifying and neglecting directions in phase space that do not lead to diverging initial conditions~\cite{Tomsovic18b}.  

Using these techniques, a complex saddle is identified with each classically allowed transport pathway.  Let us dub these ``{\it exposed saddles}''.  Amongst the infinity of exposed saddles, almost all contribute too little to be concerned with, but it is straightforward to restrict the search for real transport pathways that have sufficient amplitude to be relevant.  Nevertheless, there is still a great deal more to be done.  These works were justified intuitively and left open the question of how to identify physically relevant saddles associated with classically non-allowed transport pathways.  In contrast to exposed saddles, consider these ``{\it hidden saddles''}, in part because they are not directly discoverable using real trajectory input into a Newton-Raphson search scheme.  The focus of this work is to examine these techniques in greater detail, add additional justifications where possible, and develop a method to locate the physically relevant hidden saddles.  This extends the previous practical techniques~\cite{Pal16, Tomsovic18, Tomsovic18b} to incorporate tunneling-like phenomena and provides an interpretation of exactly what constitutes tunneling in the time domain for wave packet and coherent state propagation.

The structure of the paper is as follows, Sect.~\ref{background} defines the critical quantities, sets notations, introduces the purely quartic oscillator for simple illustrations, and discusses the background of various semiclassical approximation methods and their interrelationships.  This is followed in Sect.~\ref{justification} by a discussion of the justification of the existing methods, and develops a technique for identifying hidden saddles.  The paper concludes with a summary of the work and possible future lines of related research.

\section{Background}
\label{background}

In order to elucidate the method for implementing a search for hidden saddles and justify certain procedures, it is helpful to start with some background on the complete semiclassical approximation along with certain partial or imperfect versions of the semiclassical approximation.  As the mathematical results and manipulations can be made to appear essentially identical through the application of quadratures, it is unnecessary to treat evolution of Glauber coherent states and wave packets separately.  The discussion is presented in the language of wave packets, but it is understood that all the results carry over to coherent states.  As a final note, for simplicity the discussion and equations given here are reduced to their single degree of freedom forms.  Although, the interest is in multi-degree-of-freedom systems, it is much easier to illuminate the basic ideas with a simple example and the equations reduced to their one degree of freedom forms.  All the necessary multi-degree-of-freedom equations can be found elsewhere, for example in~\cite{Pal16,Tomsovic18b}, and the additional ideas presented here extend to the many degrees of freedom case.

\subsection{Coherent states}
\label{cs}

A Glauber coherent state describing a bosonic many-body system takes the normalized form
\begin{equation}
\label{cse}
| z \rangle = \exp \left(-\frac{\left| z \right|^2}{2} + z \hat a^\dagger \right)| 0\rangle
\end{equation}
with the properties that
\begin{equation}
\hat a | z \rangle = z | z \rangle\ , \quad \left[ \hat a, \hat a^\dagger\right] = \mathbb{1}\ , \quad \langle z | \hat a^\dagger \hat a | z \rangle = |z|^2 \ .
\end{equation}
Hence, a coherent state's mean number of bosons $n = \langle z | \hat n | z \rangle = |z|^2$.  Its evolved form $|z(t)\rangle = U(t) | z \rangle$ does not remain a coherent state, but the overlap with the bra vector version of another coherent state $\langle z^\prime|$ can be viewed as a matrix element of a coherent state path integral, $\langle z^\prime | U(t) | z \rangle$.  In the language of wave packets, this is often termed a correlation function or transport coefficient.

Introducing the quadrature (Hermitian) operators normalized as follows
\begin{eqnarray}
\hat p & = & i \sqrt{\frac{\tilde \hbar}{2}}\left( \hat a^\dagger - \hat a \right) \nonumber \\
\hat x & = &  \sqrt{\frac{\tilde \hbar}{2}}\left( \hat a^\dagger + \hat a \right) \nonumber \\
\end{eqnarray}
the operator pair behaves mathematically exactly like the quantized version of a canonical pair of momentum and position variables, even though they cannot be interpreted as some boson's momentum and position.  Setting $ \langle z | \hat x | z \rangle =  q_\alpha$ and  $\langle z | \hat p | z \rangle = p_\alpha$,  then $z =(q_\alpha + i p_\alpha)/\sqrt{2\tilde \hbar}$.  The $\tilde \hbar$ constant is written to look like the usual $\hbar$ to emphasize the analogy, but is rather usefully set to the inverse number of bosons making the $n\rightarrow \infty$ limit equivalent to the $\tilde \hbar \rightarrow 0$ limit.  With the quadrature operator $\hat x$'s eigenbasis representation $\hat x | x \rangle = x | x \rangle$, the coherent state takes the form
\begin{align}
\label{cohquad}
\langle x | z \rangle =  \left(\frac{1}{\pi \tilde \hbar}\right)^{1/4} & \exp\left[ - \frac{\left(\ x - q_\alpha \right)^2}{2\tilde \hbar}  + \frac{i p_\alpha}{\tilde \hbar} \left(x - q_\alpha \right) \right. \nonumber \\ 
& \qquad \qquad \qquad \left. + \frac{i}{2 \tilde \hbar}p_\alpha q_\alpha \right] \ .
\end{align}
Thus, an appropriate projection of the coherent state results in a Gaussian wave packet form~\cite{Glauber63} with parameters which can be straightforwardly mapped onto those of a wave packet.  Compared with Eq.~\eqref{wavepacket} just ahead, the forms are equivalent with the replacements $\tilde \hbar \rightarrow \hbar$ and $1 \rightarrow b_\alpha$.  One exposition of the parameter mapping is given in Appendix A of~\cite{Tomsovic18b}.  There are sufficient wave packet parameters in Eq.~\eqref{wavepacket} to squeeze the coherent states and rotate the quadrature operators.  The focus now turns to Gaussian wave packets.

\subsection{Gaussian wave packets}
\label{wp}

A normalized Gaussian wave packet may be parameterized as follows:
\begin{align}
\label{wavepacket}
\phi_\alpha(x) = & \exp\left[ -  \frac{ b_\alpha}{2\hbar}  \left(\ x - q_\alpha \right)^2 +\frac{i p_\alpha}{\hbar} \left(x - q_\alpha \right) + \frac{i}{2\hbar}p_\alpha q_\alpha \right] \nonumber \\ 
& \qquad \times \left[\frac{ b_\alpha+ b^*_\alpha}{2\pi\hbar}\right]^{1/4},
\end{align}
where the subscript $\alpha$ is a label for the parameters that define the particular wave packet, $x$ is the position variable for the quantum system, and $(q,p)$ are the canonically conjugate position-momentum phase space variables for the analogous classical system.  The real centroid is given by $(q_\alpha, p_\alpha)$ and the width by $b_\alpha$ (if $b_\alpha$ is complex, the wave packet has a chirped phase dependence).  All the calculations in this paper, use
\begin{equation}
\label{par}
(q_\alpha,p_\alpha)=(0,20),\: b_\alpha=32 
\end{equation}
as a convenient set for illustration ahead.  This form has the advantage that $\hbar$ does not explicitly appear in the equations for the two-point boundary value problem given ahead; i.e.~the equations become purely classical.  Note that this wave packet form is given here so as to have the exact same global phase convention as the coherent state of Eq.~\eqref{cse}, but the global phase term could just be dropped.  It also leaves the overall shape of its Wigner transform independent of $\hbar$, other than the volume (overall scale).  Similarly to a coherent state, for nonlinear dynamical systems the evolved wave packet $\phi_\alpha(x;t)$ is generally not Gaussian.  Note that a matrix element of the coherent state path integral could be expressed using quadratures analogously to 
\begin{equation}
\label{ampc}
{\cal A}_{\beta\alpha}(t) = \int_{-\infty}^\infty {\rm d}x\ \phi^*_\beta( x) \phi_\alpha( x;t)
\end{equation}
for wave packets, and which could be thought of as a transport coefficient.  If $\beta=\alpha$, it would be a diagonal element or a return amplitude.

\subsubsection{Lagrangian manifolds}
\label{lmp}

Lagrangian manifolds play a central role in semiclassical approximations, i.e.~WKB theory~\cite{Maslov81}.  They provide a very geometric picture of the application of the saddle point approximation.  Using the parameterization of Eq.~\eqref{wavepacket}, the appropriate manifold for a Gaussian wave packet was identified in~\cite{Huber88} as
\begin{equation}
\label{constraints1}
b_\alpha \left(  {\cal q} -  q_\alpha\right) + i \left(  {\cal p} -  p_\alpha\right) = 0,
\end{equation}
where $({\cal p}, {\cal q})$ are chosen from the sets of all complex positions and conjugate momenta satisfying this equation.  This generates a complex line in a two dimensional complex phase space (complex position and momentum), somewhat akin to a plane embedded in a four dimensional space.  Likewise for the complex conjugate wave packet (dual version), the equation is
\begin{equation}
\label{constraints2}
b^*_\alpha \left(  {\cal q} -  q_\alpha\right) - i \left(  {\cal p} -  p_\alpha\right) = 0.
\end{equation}
For a position ket or bra vector, the Lagrangian manifold would be the value of the position and the set of all complex momenta.

One consequence of the Lagrangian manifold being necessarily complex is that it blurs the distinction between classically allowed and non-allowed processes.  In ordinary WKB, the tori used for constructing wave functions are real Lagrangian manifolds, the intersections of two manifolds generate stationary phase points (not saddles), and the exponentially decaying tails of wave functions require propagation with imaginary time or some analytic continuation, for example, the introduction of imaginary momenta~\cite{Creagh98}.  Here it is the hidden saddles which correspond to classically non-allowed processes. Since both hidden and exposed saddles are linked to complex trajectories, ``real'' versus ``complex'' is not the distinguishing factor.  There is however the intuitive notion of complex phase points close or near to being real; see~\cite{Vanvoorhis02} and references therein.  Since distance is not defined in phase spaces, a priori, the concept of ``near'' is not well defined.  Even worse, a saddle can have an arbitrarily large imaginary component in its initial condition and still be exposed.  Part of our work here is to add some precision to this concept, which ends up helping to categorize the saddles into exposed or hidden groups.   Just as the distinction of real and complex trajectories is physically significant in the case of ordinary WKB theory, so is the exposed/hidden saddle distinction important physically.  It also alters considerably how one must search for the saddles, which is treated ahead.

\subsubsection{Two-point boundary value problem}
\label{tpbvp}

The two-point boundary value problem that arises in the saddle point approximation can be succinctly expressed using these manifolds.  In essence, the manifold corresponding to the ket vector is propagated to time $t$ according to the complexified classical dynamical equations of motion, and its intersections with the manifold associated with bra vector give the trajectories that define the saddle points' properties.  Letting  $( {\cal p}_0, {\cal q}_0)$ represent the complex initial condition for a trajectory and $({\cal p}_t,{\cal q}_t)$ the complex phase point that is generated by evolving this initial condition to time $t$, the complete two-point boundary value problem can be written as
\begin{eqnarray}
\label{sadcond1}
 b_\alpha \left( {\cal q}_0 - q_\alpha\right) + i \left(  {\cal p}_0 - p_\alpha\right) &=& 0, \nonumber \\
b^*_\beta \left( {\cal q}_t - q_\beta\right) - i \left( {\cal p}_t - p_\beta\right) &=& 0.
\end{eqnarray}
Each solution of these equations is a potential saddle point in the theory for transport coefficients.  If instead, one is interested in the evolution of the wave packet itself, the equations would be given by
\begin{eqnarray}
\label{sadcond2}
 b_\alpha \left( {\cal q}_0 - q_\alpha\right) + i \left(  {\cal p}_0 - p_\alpha\right) &=& 0, \nonumber \\
{\cal q}_t - x &=& 0,
\end{eqnarray}
where $x$ is the position argument of the evolved wave packet (a momentum representation is also possible, but not used here).  For either boundary value problem, nonlinear dynamical systems lead to an infinite number of solutions to these equations.  Almost all of them must be thrown away due to boundary conditions or due to their contributions being vastly smaller than the errors inherent in the semiclassical approximation.  Excluding them from the search algorithm by design is highly desirable and the basis of the works~~\cite{Pal16,Tomsovic18b}.

\subsection{Exposed and hidden saddles}
\label{explainsaddles}

The solutions of Eqs.~\eqref{sadcond1} and \eqref{sadcond2} contain as a subset all the necessary (i.e.~physically relevant) saddle points of the corresponding integrals.  On the other hand, taken alone as an abstract set of equations there is no indication at face value which of the infinity of solutions must be thrown away, which must be kept, which can be related to real classical transport processes, which cannot, etc... A priori, there is no interpretation of tunneling processes in the coherent state/wave packet propagation as all solutions are complex.  From a mathematical perspective, it is neither trivial to locate the solutions nor unpack how to separate out the relevant saddles and determine what their physical interpretation is.  For example, excluding simple standard illustrations of Stokes phenomena, such as the Airy function and the construction of its Stokes and anti-Stokes lines, it is highly desirable to avoid those lines' explicit construction and use, especially as a function of many complex variables.  The interest is in finding a direct, practical approach to these challenges.

One procedure recently developed~\cite{Pal16, Tomsovic18b} is to identify exposed saddles directly using a Newton-Raphson scheme.  Imagine to begin with, running all possible real trajectory initial conditions, $(q_0, p_0)$, and letting them serve as the initial input to the Newton-Raphson search algorithm.  Excluding  cases in which convergence is impacted by singular trajectories~\cite{Huber88} and complex time paths are needed (see the ``bald spot'' problem~\cite{Petersen14}), certain groupings of initial conditions lead to the exact same saddle point because they are in the same basin of attraction.  Thus, the real phase points can be grouped into subregions (with negligible exceptions).  All trajectories within a single subregion lead to the same saddle point.  A subregion's trajectories all shadow each other and behave similarly in a dynamical sense.  Each subregion thus represents a single classical transport pathway, and it is necessary only to find a single reference trajectory to represent that pathway as an input for the actual Newton-Raphson search.  This implies that a complex exposed saddle trajectory shadows the real trajectories associated with that pathway.  In particular, the stability matrix relation
\begin{equation}
\left( \begin{array}{c} \delta \cal p_t \\ \delta \cal q_t \end{array} \right) =  \left( \begin{array}{c} \bf{M_{11}} \\ \bf{M_{21}} \end{array} \begin{array}{c} \bf{M_{12}} \\ \bf{M_{22}} \end{array} \right)
\left( \begin{array}{c} \delta \cal p_0 \\ \delta \cal q_0 \end{array} \right) 
\label{delta}
\end{equation} 
allowing for complex initial condition deviations from the real reference trajectory would adequately describe the true exposed saddle trajectory.  The exposed saddle acquires the interpretation that it accounts for the contributions of the classically allowed transport pathway to which the reference trajectory belongs, and there is a one-to-one correspondence within the collection of exposed saddles and transport pathways.  

An immediate consequence of this shadowing relationship is that the exposed saddle's classical action would almost exclusively vary in its real part as a function of time or other parameters leading to predominantly oscillatory contributions, say for example, to the propagated wave function.  Its imaginary part would be positive, constrained to vary slowly, and account mainly for the Gaussian weighting of the initial condition and/or final conditions in the wave packets' Wigner transform.

The above procedure necessarily leaves out the subset of physically relevant saddles associated with tunneling and/or classically non-allowed processes, which we classify as hidden saddles since they are hidden from the Newton-Raphson search just described.  One main purpose of this paper is to extend that procedure so as to obtain these saddles as well; see Sect.~\ref{sec:hs}.  Evidently, these saddle trajectories behave rather differently, and cannot shadow any of the real trajectories.  One may anticipate that their classical actions vary as a function of parameters more predominantly in their positive imaginary parts rather than their real parts.  In this way, they contribute largely to exponentially decaying behaviors.

The complexification of the classical dynamics does have some less familiar fundamental impacts.  An example is given in~\cite{Petersen14} of saddles obscured by bald spots.  The basic idea is that singular complex trajectories create dynamical branch cuts and to arrive at some saddle solutions, it is necessary to deform a real time propagation path into some complex time domain path.  Thus, one can also anticipate the complication that there exists another class of saddles, those behind bald spots, which require complex time integration paths to follow.  This class does not exist for the simple example described next nor was it significant enough in any of the Bose-Hubbard model calculations of~\cite{Tomsovic18, Tomsovic18b} to impact the accuracy.  They will not be considered further here.
 
\subsection{The purely quartic oscillator}
\label{quarticoscillator}

The one degree of freedom purely quartic oscillator has a number of simple features that makes it ideal for illustrating the ideas discussed in this paper.  Its analytically continued form [complex (${\cal p}, {\cal q}$)] is given by
\begin{equation}
\label{quosc}
H( {\cal p}, {\cal q}) = \frac{{\cal p}^2}{2} + \lambda {\cal q}^4,
\end{equation}
where $\hbar=m=1$ and $\lambda= 0.05$ are the values taken for all the illustrations shown in this paper.  The corresponding Schr\"odinger equation is given by
\begin{equation}
 i\frac{\partial}{\partial t} \phi(x;t) = \left(-\frac{\partial^2}{2\partial x^2} + \lambda x^4 \right) \phi(x;t).
\label{schroedinger}
\end{equation}
Being a homogeneous Hamiltonian, the real classical dynamics of this system leads to simple scaling relations amongst the quantities $(q,p;t;E)$.  For example, propagating an initial condition $(q_{E_0},p_{E_0})$ for a time $t$ that belongs on the $E_0$ energy surface leads to a scaled trajectory replica on any desired second energy surface $E$ according to:  
\begin{align}
\label{scaling}
q_E(t/\gamma ) & = \gamma q_{E_0}(t), \nonumber \\
p_E(t/\gamma ) & = \gamma^2 p_{E_0}(t), 
\end{align}
where $\gamma=(E/E_0)^{1/4}$.  Classical actions, defined as $S = \int {\cal L} {\rm d}t$, scale as $\gamma^3$, and the period of a closed orbit, $\tau_{E_0}$, scales as $\tau_{E} = \tau_{E_0}/\gamma$.  Therefore, if $\gamma = n^{\pm 1}$, where $n$ is any positive integer, the two periodic orbits  on different energy surfaces would have periods that are an integer multiple of each other.  In the time that the longer period orbit closes once, the shorter period orbit retraces itself $n$ times.

\begin{figure}
    \begin{subfigure}
            \centering
            \includegraphics[width=8 cm]{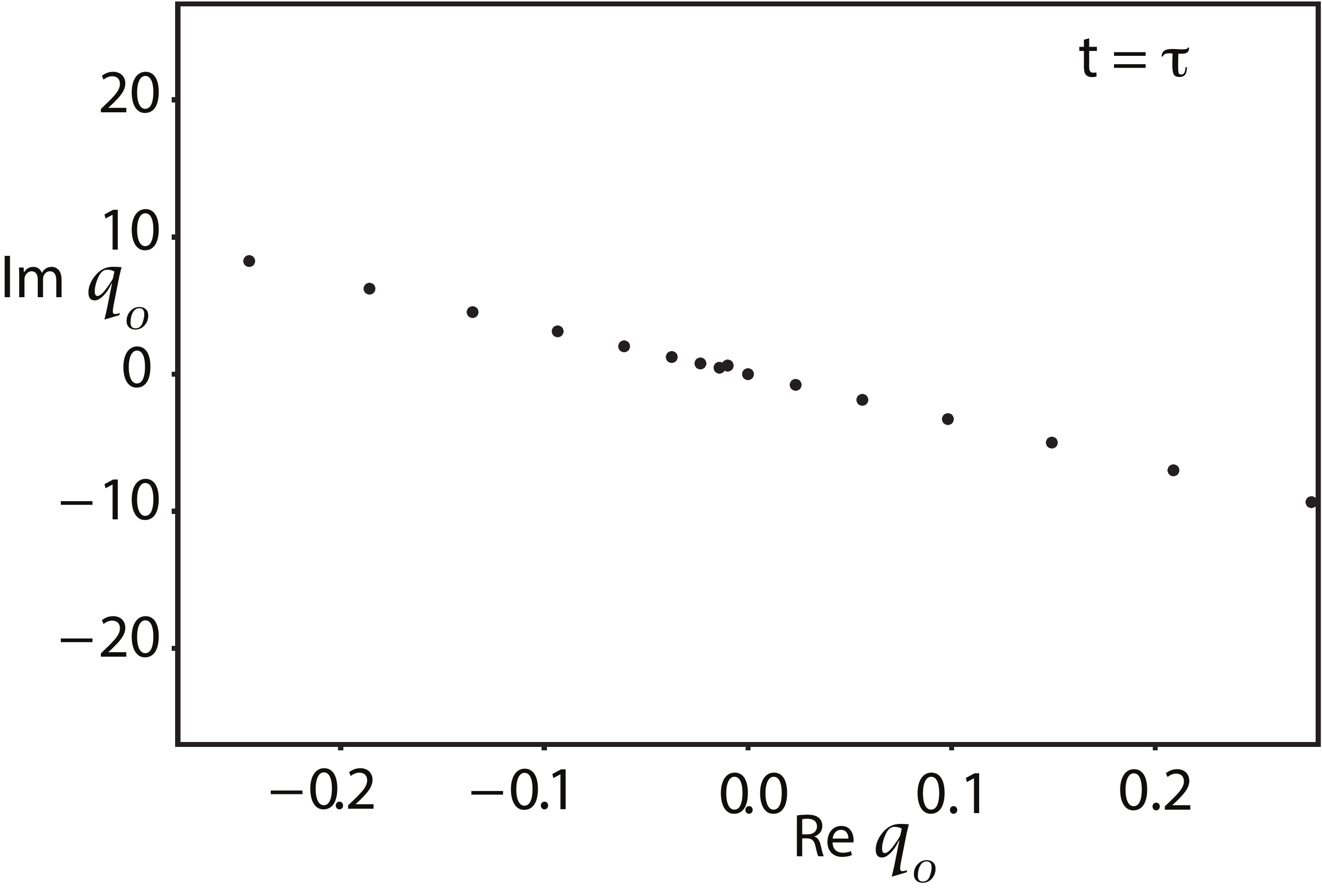}
            \label{densitya}
    \end{subfigure}
    \begin{subfigure}
            \centering
            \includegraphics[width=8 cm]{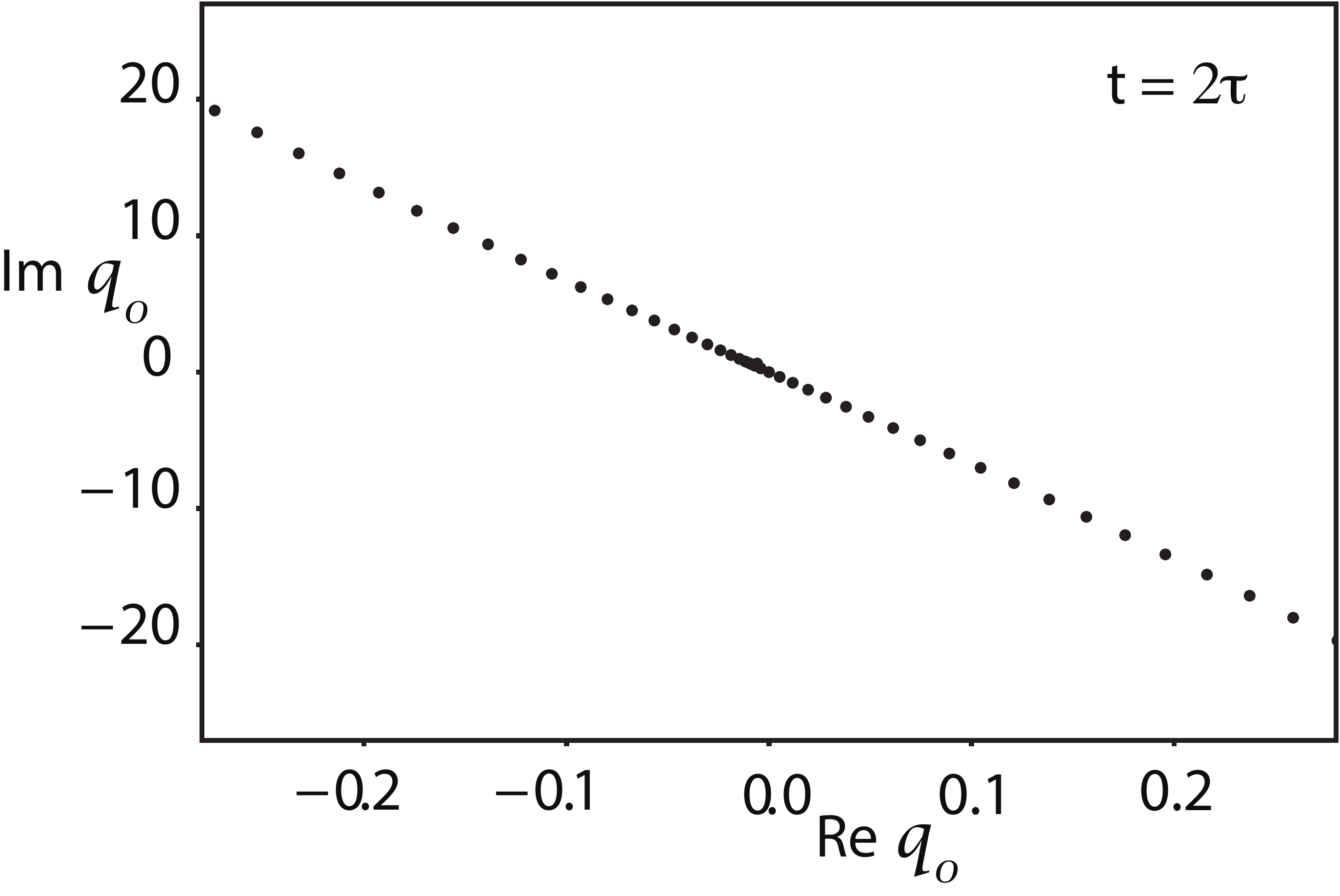}
            \label{densityb}
    \end{subfigure} 
    \caption{Initial conditions of exposed saddles satisfying Eq.~\eqref{sadcond2} for a wave packet defined by Eqs.~\eqref{wavepacket} and~\eqref{par}, and propagated with the Hamiltonian, Eq.~\eqref{quosc}.  The complex momentum of a saddle's initial conditions follows from the complex position using Eq.~\eqref{constraints1}.  In the upper panel, the propagation time is the period of the central trajectory with initial condition $(q_0,p_0)= (0,20)$ and in the lower panel twice this period.  The density of exposed saddles increases with increasing time, but even in the limit of small times, the total number of just the exposed saddles is infinite.  The only exception is for $t=0$. \label{density}}
\end{figure}

The potential generates nonlinear dynamical equations that lead to sufficiently complicated behaviors for our purposes.  As a first example, see Fig.~\ref{density}, which illustrates the placement of just the exposed saddle initial conditions in a plane for the evolution of an initial wave packet evaluated at $x=0$ and some fixed time $t$.
There is no limiting domain within the Lagrangian manifold for the solutions, and an infinity of solutions is implied over the manifold's infinite domain.  In particular, note that these exposed saddles can have imaginary parts of their initial conditions that extend out to $\pm$ infinity.  What alters with propagation time is the density of solutions.  As time increases, double the propagation time is shown in the lower panel, the density increases.   At $t=0$, there is effectively only one saddle left with all the other saddles pushed out to infinity.  Keep in mind that these exposed saddles are only a small subset of all the saddles.  There are solutions everywhere throughout the complex ${\cal q}$ plane, not just approximately along a line, which is where the exposed saddles find themselves.

\subsection{Wigner transform}
\label{wt}

The Wigner transform of a wave packet (ket-like) generates a multivariate Gaussian function, which can be thought of as a density of real classical initial conditions that underlie a quantum wave packet and account for the uncertainty principle.  It can be used to help understand the partial semiclassical approximation known as linearized wave packet dynamics~\cite{Heller75}, and is essential for discussing the basis of an enhanced approximation known as an off-center, real trajectory method~\cite{Tomsovic91b,Tomsovic93,Tomsovic93b,Barnes93,Barnes94}.

The Wigner transform of a wave packet parametrized as in Eq.~\eqref{wavepacket} is given by
\begin{equation}
{\cal W}( p, q)= \frac{1}{\pi \hbar} \exp \left[ - \left(p - p_\alpha, q - q_\alpha \right) \cdot \frac{{\bf A}_\alpha}{\hbar} \cdot \left(\begin{array}{c} p - p_\alpha \\ q - q_\alpha \end{array}\right) \right], 
\end{equation}
where ${\bf A}_\alpha$ is
\begin{equation}
\label{mvg}
{\bf A}_\alpha = \left(\begin{array}{cc}
1/c & d/c  \\
d/c  & c +  d^2/c  \end{array}  
\right),   \qquad {\rm Det}\left[ {\bf A}_\alpha \right] =1
\end{equation}
with the association 
\begin{equation}
\label{mvgwf}
b_\alpha =c + i d.
\end{equation}
The $2 \times 2$ dimensional matrix ${\bf A}_\alpha$ is real and symmetric.  If $b_\alpha$ is real, there are no correlations between $p$ and $ q$ ($d$ vanishes); i.e.~the wave packet is not chirped.  The off-diagonal elements (blocks in more degrees of freedom) of  ${\bf A}_\alpha$ disappear.

\subsection{Partial semiclassical theories}
\label{pcd}

There are two important partial semiclassical approximations known as linearized wave packet dynamics and off-center, real trajectory methods, respectively, in order of increasing sophistication.  The former linearizes the dynamics completely, and the latter contains the nonlinear dynamical information, which is much, much closer to GGWPD. 

\subsubsection{Linearized wave packet dynamics (LWPD)} 

The main idea involves solving the equations of motion for the parameters that define an evolved Gaussian, including the center,  width, and phase/normalization factor in a global fashion.   Its validity is constrained to cases where the quadratically expanded potential around the central trajectory is a good approximation to the dynamics within the phase space volume defined by the Wigner transform of the evolving wave packet.  As the wave packet spreads out with increasing time, it soon extends well beyond the domain of validity for the quadratic expansion, which leads to the breakdown of the approximation.

\begin{figure}
   \centering
   \includegraphics[width=8 cm]{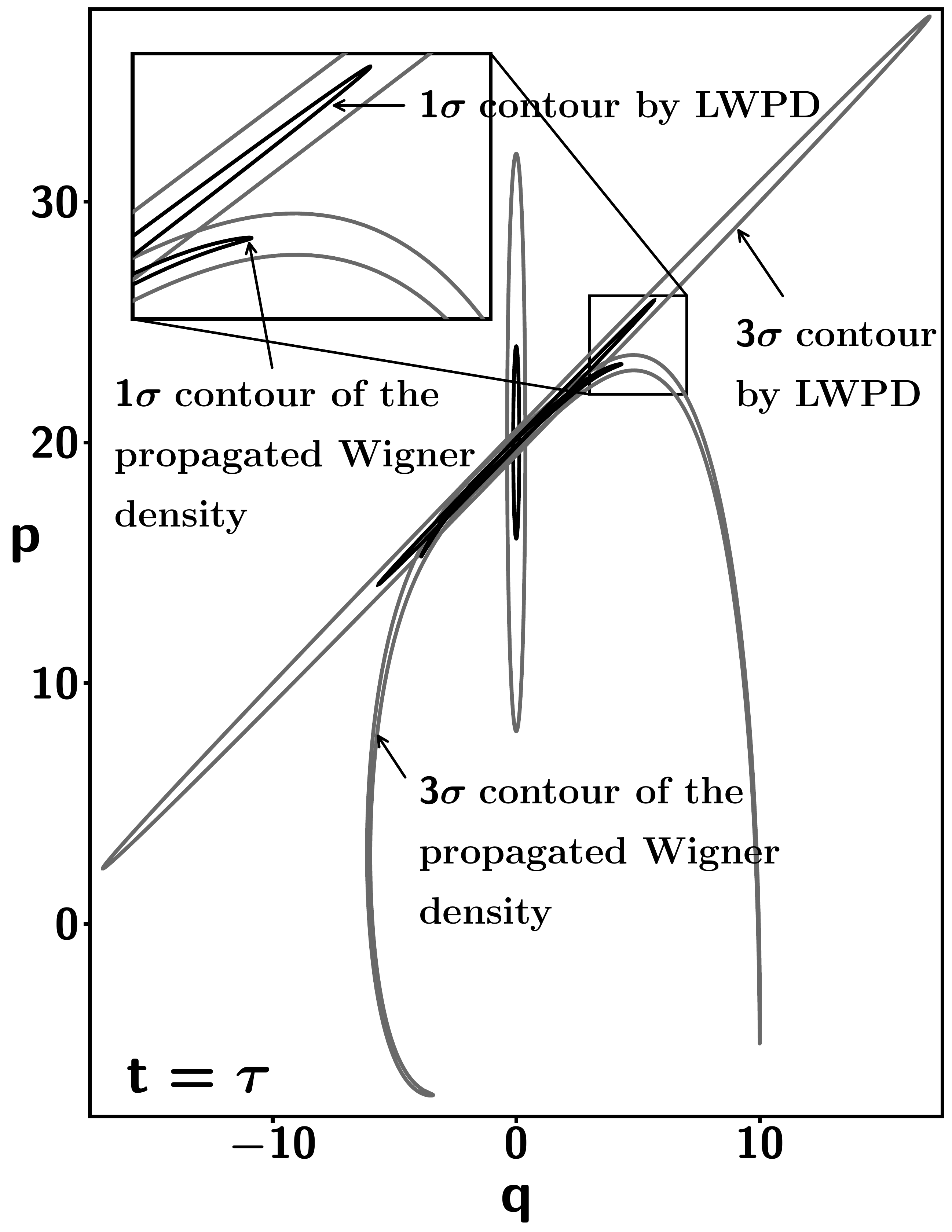}
   \caption{Illustration of the limitations of LWPD. The vertically oriented ellipses are contours of the  initial wave packet's Wigner transform using the parameters of Eq.~\eqref{par}. This density is classically propagated for $1$ period of the central trajectory's motion ($\tau$), forming curvy foliations. The LWPD produced wave packet generates stretched and rotated ellipses.  Only phase points within $1\sigma$ are still approximately linearly related to the central trajectory by this propagation time.}
 \label{lwpd}
\end{figure}

The validity is limited to pre-Ehrenfest time scales.  One method for determining this time scale in practice is to consider two evolved densities.  The first is the initial wave packet's Wigner transform evolved classically (numerically), and the second is the Wigner transform of the LWPD evolved wave packet.  Let the overlap integral of these two densities at $t=0$ be normalized to unity.  Then LWPD remains an accurate approximation until a time such that the overlap drops precipitously from one (this is the Ehrenfest time scale).  This is because LWPD has only linearized dynamical information about the central trajectory whereas the density of initial conditions incorporates nonlinear dynamical effects relative to the central trajectory.  See Fig.~\ref{lwpd} for an illustration.  The one and three standard deviation ($\sigma$) ellipses of LWPD appear straight and are superposed on top of the $1$ and $3 \sigma$ curves of the initial state's propagated Wigner transform.  After one period of the central trajectory's motion some information about the wave packet central region may still be given correctly by LWPD, but beyond the $1\sigma$ curve, the dynamical information is no longer accurate.  This illustrates the breakdown of LWPD at short time scales for nonlinear dynamical systems.

\subsubsection {Off-center, real trajectory method}

The important idea underlying this method is that for any nonlinear dynamical system over time a localized density of initial conditions disperses and especially in a bounded system, the dispersed evolved density must fold over or one might say create foliations.  In some local region of phase space, these foliations slice through (with increasing density as time increases), and each one contains an infinite collection of nearly identically behaving trajectories; i.e.~a single classical transport pathway.  See the study of the production of whorls and tendrils in~\cite{Berry79b}.  In other words, the stability matrix of one member of the local set predicts rather well the behavior of its neighbors within the foliation.  On the other hand, each foliation has a completely different dynamical character and represents a quite distinct transport pathway.  By identifying a member from each foliation, which necessarily involves off-center initial conditions [those initial conditions within the Wigner density, but not equal to $(q_\alpha,p_\alpha)$], and following the essential prescription of LWPD, a wave is constructed whose information is locally correct.  Summing over the contributions of all the foliations incorporates the nonlinear dynamical features.  This technique locally reconstructs the features of the evolving wave packet including quantum interference between transport pathways, and its validity goes well beyond the Ehrenfest time scale~\cite{Tomsovic91b,Oconnor92,Sepulveda92}.
\begin{figure}
         \centering
         \includegraphics[width=8.0 cm]{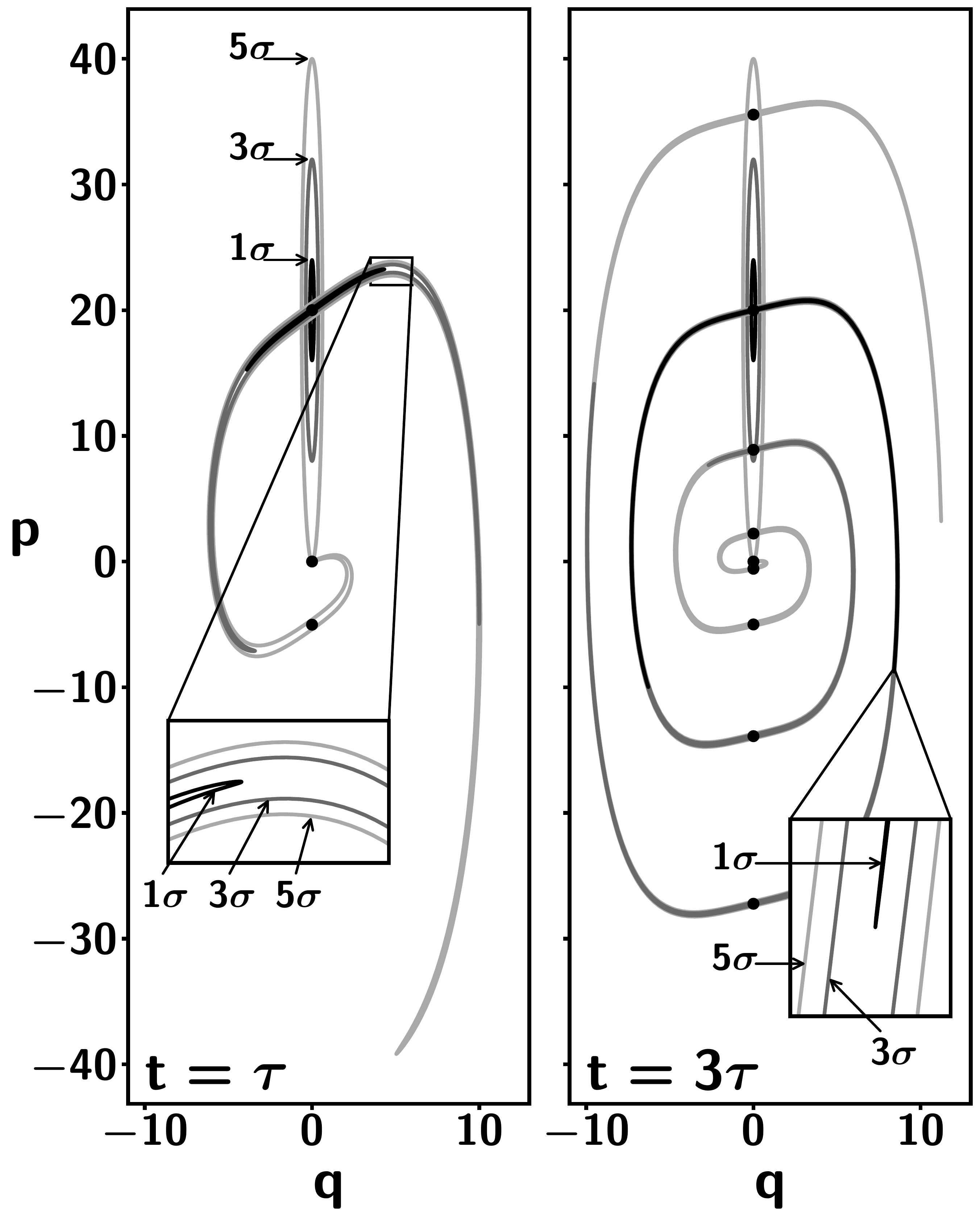}
    \caption{Illustration of distinct classical transport pathways. The vertically oriented ellipses are contours of the initial wave packet's Wigner transform using the parameters of Eq.~\eqref{par}.  The left panel is propagated classically for $1$ period of the motion ($\tau$) and the right panel $3$ periods.   They generate the distorted spirals.  The insets show expanded views of localized regions.  In the $t=3\tau$ figure, the $5\sigma$ contour can be divided into $9$ foliations (enumerated in later figures), a trajectory member of which can be used to generate initial conditions for a Newton-Raphson scheme to locate unique saddles. The blackened dots show the locations of the real parts of the $9$ associated saddles' complex final phase space points contributing to $\phi(x=0,3\tau)$ (the initial conditions follow by back propagating a time $3\tau$). An analogous statement applies to the $t=1\tau$ figure on the left.    \label{wigner}}
 \end{figure}

The Wigner transformed initial density plays a more critical role for this method since it is essentially being used to identify the foliations themselves as a function of time.  This is illustrated in Fig.~\ref{wigner} where each foliation can be used to identify a single real reference trajectory about which one can linearize the dynamics locally and that can be used to generate a contribution to the quantity of interest.  Furthermore, each of these off-center, real reference trajectories can be used in a Newton-Raphson search to locate a unique exposed saddle, which generates an important subset of the saddles of GGWPD~\cite{Pal16}.

It also lends itself to a prescription for identifying the phase space directions of fastest dispersion thereby indicating the directions that are unnecessary for developing the foliations for the quantity of interest~\cite{Tomsovic18b}.  In this way, problems with large numbers of degrees of freedom can be reduced to vastly fewer numbers.  This effective dimensional reduction makes it possible to treat fully more complicated dynamical systems.  Even if many degrees of freedom are active in this sense, they usually separate out in time scales.  Thus, one can successively add directions as time increases, but start with just one or two.  For many problems, being able to go to intermediate time ranges beyond the Ehrenfest scale is sufficient~\cite{Tomsovic18}.

\subsection{Complications due to complexifying the classical dynamics}
\label{cccd}

There are a number of complications arising if classical dynamics are analytically continued to complex phase space variables.  One of the most significant considerations is the existence of singular orbits, i.e.~orbits that acquire infinite momenta in finite propagation times~\cite{Huber88}.  If time is also allowed to be complex so that a path can be defined to avoid the singular real time point, there may exist multi-sheeted functions describing the dynamics~\cite{Ince56,Polyanin03}.  Numerically for the quartic oscillator, important dynamical quantities such as position, momentum and classical action are not multi-valued functions of time. This is consistent with it possessing the Painlev\'e property in which solutions of a differential equation have no movable branch points.  In this sense, for the pure quartic oscillator the singularities are isolated poles in a complex propagation time space. 

\begin{figure}
   \centering
   \includegraphics[width=8cm]{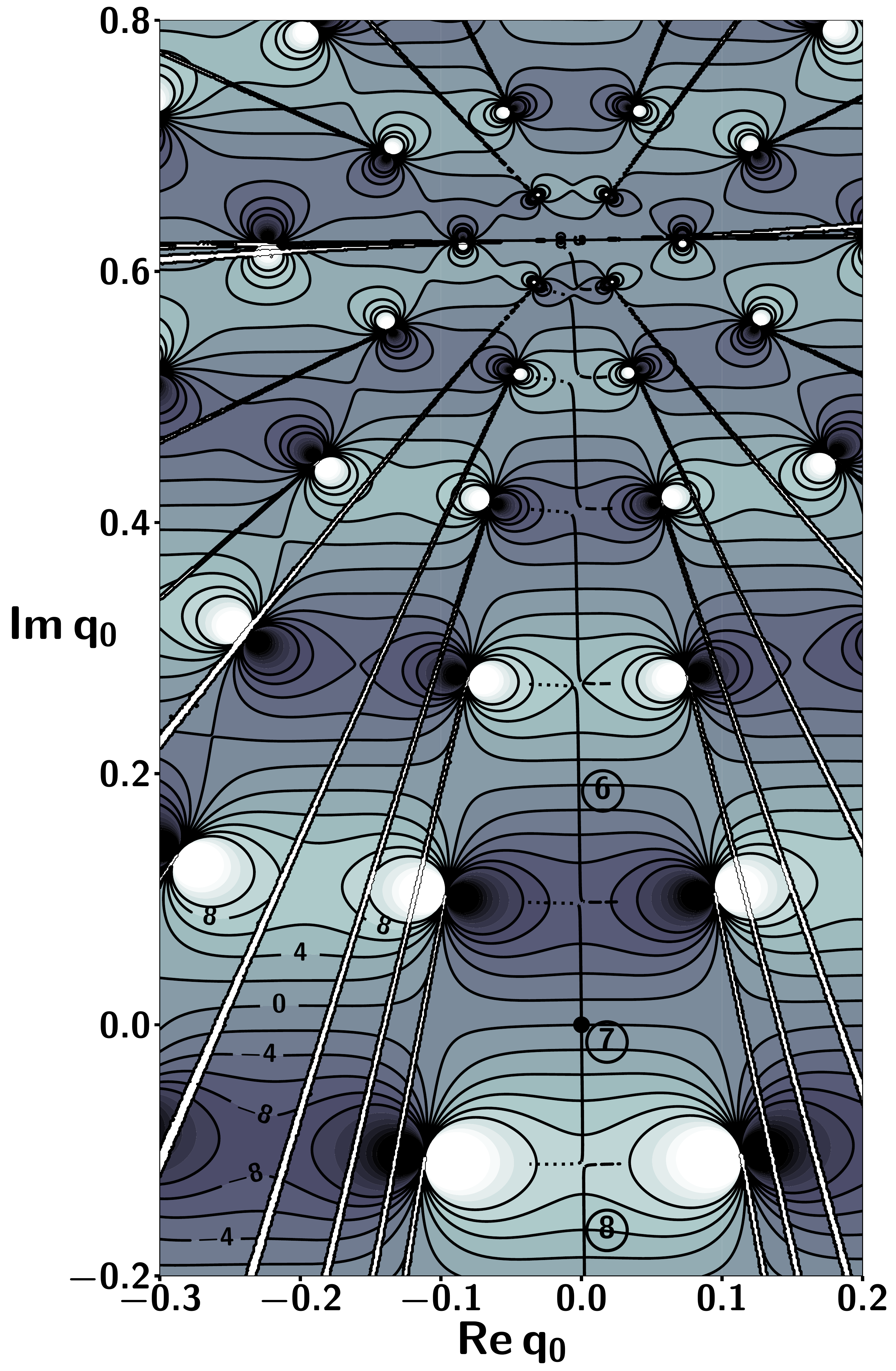}
   \caption{Contours of the final position $x$'s real part  as a function of the trajectory initial conditions on the initial wave packet's Lagrangian manifold.  Real and imaginary parts of $\cal p_0$ follow using Eq.~\eqref{constraints1} with the parameters of Eq.~\eqref{par}; a similar figure (Fig.3) was given in~\cite{Huber88}.  The black dot near \textcircled{\raisebox{-0.9pt}{7}} indicates the real initial condition of the Wigner density centroid. The initial conditions of exposed saddles associated with foliations \textcircled{\raisebox{-0.9pt}{6}}, \textcircled{\raisebox{-0.9pt}{7}}, \textcircled{\raisebox{-0.9pt}{8}} as a function of the propagated wave function's position $x$ appear as nearly vertical solid lines perpendicular to the contour lines of $Re\,x$; they are located on $Im\,x=0$ contours which are not shown.  The transition from exposed to hidden occurs at the {\it avoided crossing} of two saddles from different foliations, which occurs near the classical turning point.   The solid line turns to dashed or dotted where it transitions to hidden. The dashes are for saddles that crossed a Stokes line and must be excluded. }
   \label{contourfig}
\end{figure}

On the other hand, these singularities must come in continuous families.  Consider a trajectory which becomes singular exactly at a propagation time $t_0$.  Differentially nearby will be other trajectories that become singular at real propagation times of $t_0\pm \epsilon$.  By following the changing initial conditions for a singular trajectory continuously as the propagation time varies from $t=0$ to $t=\infty$, they can be associated to a continuously shifting set of initial conditions.  This is illustrated for the quartic oscillator in Fig.~\ref{contourfig}. 
For the real and imaginary parts of positions on the initial manifold, the real part of the final position is contoured and shown as a density plot (the complex momentum initial condition for each point in the figure can be deduced from Eq.~\eqref{constraints1}).  The trajectory calculations fail in the immediate neighborhood of singular trajectories.  This shows up as extremely narrow cones devoid of data, which   correspond to sets of initial conditions for trajectories that come too close to singular trajectories for any time up to the final propagation time of $3\tau$.  

Mathematically, the narrow cones would be lines.  All such lines appear to emanate from one of two limiting points and they extend closer to those particular points as longer propagation times are considered.  However, the precise structure of the dynamics for $\mathcal{q}$ in the neighborhood of $(0, 0.625)$ is quite complicated and difficult to discern precisely.  Modulo the difficulties in the immediate neighborhood of these points for long times, this Lagrangian manifold for the initial coherent state defined by Eq.~\eqref{wavepacket} and \eqref{par} can be partitioned into an infinity of subregions by delineating the boundary of each region with the lines of singular trajectory initial conditions.  Within any given subregion, the initial conditions can be varied continuously without encountering a singular trajectory.  In the next section, one particular subregion plays an important role.  Finally, note that for systems which do possess branch cuts (i.e.~multi-sheeted functions describing the dynamics), it would be much more difficult to delineate subregions than for the quartic oscillator.  We leave this for future study.

\subsection{Features and accuracy of the semiclassical approximation}

Even though semiclassical methods seem to be applied most often to make qualitative physical arguments, it is important to point out that their full implementation in detail leads to very quantitatively accurate approximations for systems in the appropriate limits, e.g.~short wavelength or large particle number limits.  For example, coherent state propagations in Bose-Hubbard lattices with many sites are accurately reproduced well past the Ehrenfest time scale~\cite{Tomsovic18}.  

\begin{figure}
   \centering
   \includegraphics[width=8.5 cm]{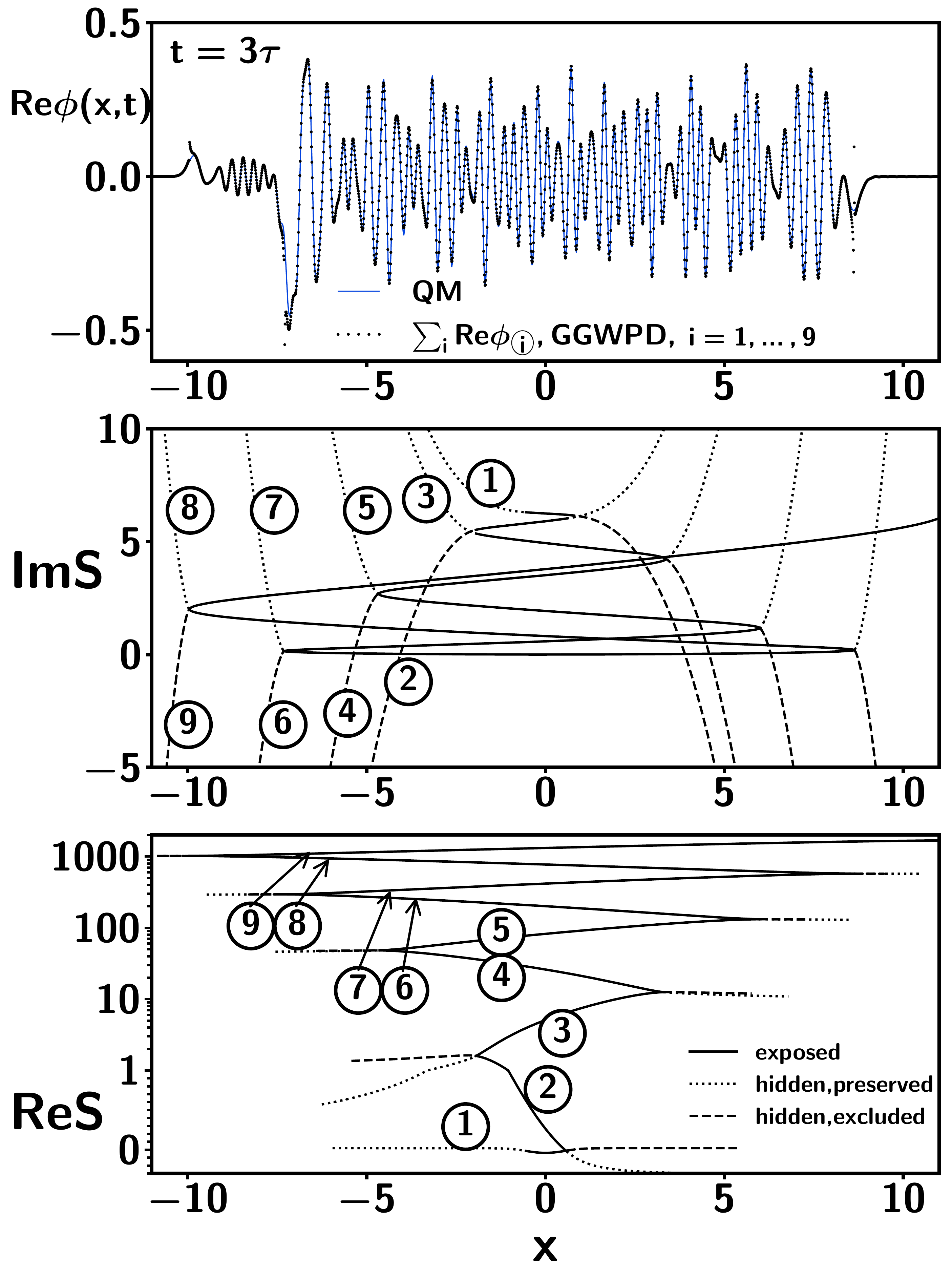}
   \caption{Comparison of the quantum propagation of a wave packet and the GGWPD approximation.  The particular wave packet illustrated is the one defined by Eq.~\eqref{wavepacket} and~\eqref{par} propagated for $3\tau$. The lower two panels show the imaginary and real parts of the classical actions, respectively.  The dashed and dotted lines indicate where saddles become hidden.  The (avoided) crossing marks the transition in how quickly the real and imaginary parts of the classical action are changing as a function of $x$, respectively.  The dashed line indicates a saddle has crossed a Stokes line, which is indicated here by a crossing of the real parts of two classical actions, and must be excluded.  The dotted lines indicate hidden saddles, which can be included, and may be physically relevant if the imaginary parts of their actions are not too large.}
 \label{ggwpd}
\end{figure}

The accuracy of the semiclassical approximation is illustrated in Fig.~\ref{ggwpd} for the much simpler quartic oscillator where it is possible to compare the full propagated wave 
function with its GGWPD approximation.  For this example, it is easier to ensure that all the physically relevant saddles have been identified.  The distorted $5\sigma$ spiral of Fig.~\ref{wigner} implies $9$ foliations, each one's boundaries approximately given by classical turning points; see Fig.~$5$ ahead in the companion paper where the foliations are labeled.  It is also straightforward to ensure that saddles on the wrong side of Stokes lines are excluded.  In Fig.~\ref{ggwpd}, the real and imaginary parts of the contributing saddles' classical actions are shown as a function of position and labeled by their respective associated foliation below the comparison.  Note that the foliation label remains valid beyond the turning point caustic positions where exposed saddles become hidden.  The transition point from exposed to hidden also marks a distinct change in the behaviors of the real and imaginary parts of the classical actions.

It turns out that all of the potentially, physically relevant exposed and hidden saddles reside within the central vertical subregion of Fig.~\ref{contourfig}.  Let's call it the {\it classical zone}.  Not a single physically relevant saddle, exposed or hidden, comes from any other subregion; we suspect that this is true more generally than just for the quartic oscillator.  A curious feature of the classical zone boundary for the quartic oscillator is that each point on it can be thought of as a limiting or accumulation point of an infinite sequence of points from an infinity of lines passing by getting ever closer to the boundary.  At least for the quartic oscillators, the saddles tend to be well away from these complicated boundary zones.  

Just for emphasis, all exposed saddles reside within the classical zone and that means an infinity, even those contributing vastly too little to be physically relevant.  Since they are exposed though, it is straightforward to implement a cutoff on their physical relevance based on the weighting of the classical reference trajectories used to locate them.  For example, the set of saddles associated with just the nine foliations arising from within the Wigner transform's $5\sigma$ contour are all within just the part of the classical zone limited by  $-0.625 \le Im(q_0)\le 0.625$ ($=5/\sqrt{2b_\alpha}=5/8$).  Any saddles from outside this restricted subregion, contribute much less than any of the dominant saddles from within.  This has the practical consequence of eliminating almost all of the Lagrangian manifold from the search technique to be used to solve the two-point boundary value problem, and therefore is a great and essential simplification.   

In the central 
\begin{figure}
   \centering
   \includegraphics[width=8 cm]{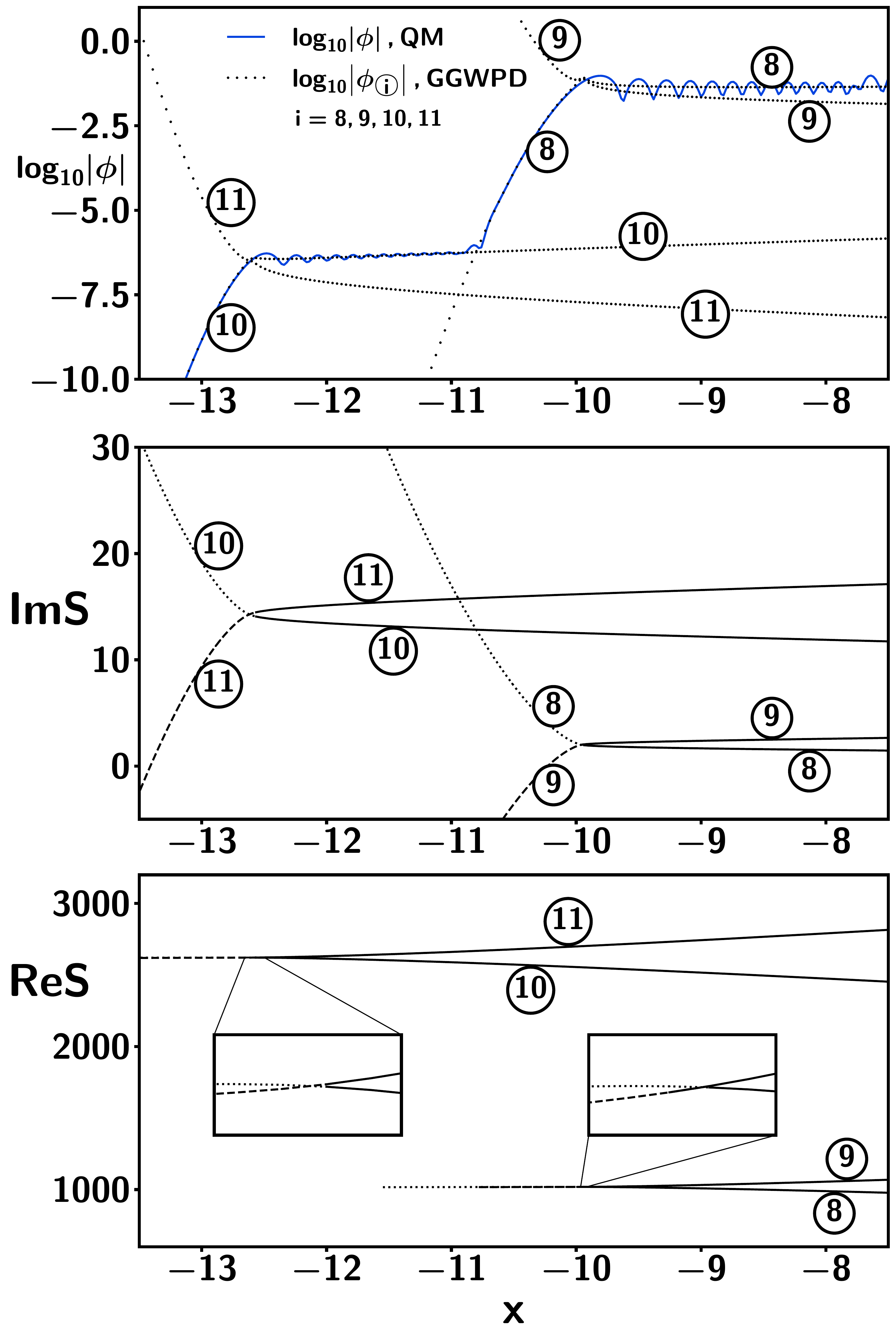}
   \caption{Expanded view of the left side of Fig.~\ref{ggwpd}.  The propagated wave function decreases quite  a bit to the left of where foliations \textcircled{\raisebox{-0.9pt}{8}}  and \textcircled{\raisebox{-0.9pt}{9}}  switch from exposed to hidden.  Just to the left of this switch, the hidden saddle \textcircled{\raisebox{-0.9pt}{8}}'s contribution is just as significant as those of the exposed saddles \textcircled{\raisebox{-0.9pt}{1\!0}}  and \textcircled{\raisebox{-0.9pt}{1\!1}}, whereas \textcircled{\raisebox{-0.9pt}{9}} has crossed a Stokes line and ceases to contribute. Foliations \textcircled{\raisebox{-0.9pt}{1\!0}} and \textcircled{\raisebox{-0.9pt}{1\!1}} are from larger-than-$5\sigma$ contours, not plotted in Fig.~\ref{wigner} and Fig.~\ref{ggwpd}.  The semiclassical inaccuracies seen where \textcircled{\raisebox{-0.9pt}{9}} and \textcircled{\raisebox{-0.9pt}{1\!1}} cross Stokes lines can be healed (uniformized) following Berry's prescription~\cite{Berry89}. }
  \label{ggwpdleft}
\end{figure}
region of the propagated wave function of Fig.~\ref{ggwpd}, the exposed saddles dominate the contributions of the hidden saddles.  Even some of the exposed saddles included are contributing with very small magnitudes.  The exposed saddles are the ones whose imaginary parts of the action are changing relatively slowly (real parts relatively quickly) as a function of position.  Also, the contribution of the hidden saddles are imperceptible there, i.e.~the saddles on the correct side of Stokes lines and possessing the opposite behavior in their real and imaginary parts as the exposed saddles' actions.  

Away from the center, the number of exposed saddles diminishes as various turning points in the real classical dynamics are surpassed.  Some of the hidden saddles begin to get comparable in magnitude to the remaining exposed saddles and even further out dominate the contributions to the wave function.  The left side of the wave function is magnified to illustrate these points in Fig.~\ref{ggwpdleft}.  In the upper panel, the logarithm of the propagated wave function magnitude is plotted.  This makes the alternation of oscillatory and exponential-like decay behaviors clearly visible.  On the right, the exposed saddles of foliations \textcircled{\raisebox{-0.9pt}{8}}  and \textcircled{\raisebox{-0.9pt}{9}}  are the dominant contributions and they create an oscillatory ``plateau'' in which the overall magnitude is barely changing.  Just beyond $x=-10$, they pass a caustic where they both convert from exposed to hidden.  Foliation \textcircled{\raisebox{-0.9pt}{9}}'s saddle contribution must be thrown away (Stokes phenomenon) and only foliation \textcircled{\raisebox{-0.9pt}{8}}'s saddle contribution kept.  This creates a ``shoulder'' of exponential-like decay because its imaginary action is changing rapidly, but not its real part.  Its contribution dominates up to the point where its decay renders it smaller than the magnitude of the next plateau created by the exposed saddles \textcircled{\raisebox{-0.9pt}{1\!0}}  and \textcircled{\raisebox{-0.9pt}{1\!1}}.  This circumstance repeats itself further to the left where \textcircled{\raisebox{-0.9pt}{1\!0}}  and \textcircled{\raisebox{-0.9pt}{1\!1}} pass a caustic and convert to being hidden.  This exponential-like decaying behavior cannot be described properly based only on real classical transport processes.  Thus, the hidden saddles are responsible for properly describing tunneling effects in the evolution of wave packets.

 This point where two saddles nearly coalesce and convert from being exposed to hidden can be defined more precisely by where the real parts of their actions cross~\cite{Huber88}.  Unlike in ordinary WKB where the coalescence is perfect at classical turning point caustics, for coherent states and wave packets generally the real and imaginary parts of the classical actions do not cross at the same point and in that sense these are just near coalescences that appear as avoided crossings.  Even the small inaccuracies due to discontinuously dropping a saddle can be improved upon following a uniformization prescription due to Berry~\cite{Berry89}.  The upshot is that GGWPD carried out in full is highly accurate for the entire propagated wave function including the exponentially decaying tails, and even the most troublesome locations near caustics can be improved through uniformization techniques.

\section{Justifications for implementation strategies and hidden saddles}
\label{justification}

The focus of this section is justifications of off-center, real trajectory methods~\cite{Tomsovic93,Barnes94},  implementations of GGWPD~\cite{Vanvoorhis02,Vanvoorhis03,Pal16,Tomsovic18}, and incorporating hidden saddles.  These techniques have used reasonable, physically motivated, but heuristic arguments and numerical evidence to support their developments.  However, certain concepts such as the already mentioned idea of a complex phase space point being {\it near}~\cite{Vanvoorhis02} a real one are somewhat vague or misleading.  It is possible to add some precision to such ideas.

\subsection{Nearness of complex to real trajectories}
\label{nearness}

We have thought of three somewhat related possible avenues of adding precision to the concept of {\it nearness} between a complex and real trajectory.  

\begin{itemize}

\item[1.] The first route is to insist that ``nearby'' means a Newton-Raphson search beginning from a real reference trajectory taken from some particular foliation generated by the dynamics converges to a unique complex saddle trajectory that can consequently be associated with that foliation.  After all, with some small technical modifications, this is how the exposed saddles are identified~\cite{Vanvoorhis02, Pal16, Tomsovic18b}.  The complex saddle trajectories do not have to possess tiny imaginary parts for this to happen, but they do have to shadow their real trajectory counterparts in the sense that the stability matrix relation, Eq.~\eqref{delta}, holds between the saddle and reference trajectories.

\item[2.] A second route, which we find a bit more compelling, is related to how the saddle point approximation is carried out.  It involves a locally quadratic expansion of some very complicated action function about a point for which the linear term vanishes.  However, inside a domain in which this function is behaving quadratically enough, an expansion about any point inside that domain is possible to describe the behavior of that function.  For the action functions of dynamical systems these domains tend to be highly asymmetric.  One direction in the domain may be very "compressed", whereas another is quite "elongated", which, of course, complicates the indeterminate idea of ``near''.  There are exposed saddles with imaginary parts of their initial conditions approaching infinity as well.  Nevertheless, the quadratic expansion is used as an approximation locally for the purposes of integration (usually with the limits extended to $\pm \infty$).   The only distinction between the two approaches is that, except for the expansion made exactly at the saddle point, there exists a linear term.  Incorporating the linear term into the Gaussian integral being performed would lead to the exact same value for a perfectly quadratic function, and gives nearly the same result if the function is sufficiently quadratic locally.  

Therefore, a second definition of the complex saddle trajectory being near a real trajectory is that the Gaussian saddle integration gives the same result as the Gaussian integration using the related neighboring real reference trajectory as an expansion point and incorporating the non-vanishing linear term.  Adopting this definition, the off-center real reference trajectory method is partly a poor man's GGWPD in which one does not bother with performing the Newton-Raphson search to find the true saddle point, but each contribution from the Gaussian integrals expanded about a real trajectory gives essentially the same result as the Gaussian integral performed at the real trajectory's associated saddle point.  It is important to note that this does not imply that the complex saddle is near a real point.  It can have very large imaginary components of position and momentum.  If a real classical transport pathway gives an extremely small contribution, then the complex saddle necessarily has a significant imaginary part of the complex classical action attenuating its contribution equivalently.  It is also true that this necessarily excludes hidden saddles from the method.  Therefore, one expects accurate approximations using the techniques of~\cite{Oconnor92,Tomsovic93,Barnes94} where the quantum dynamics is not dominated by hidden saddle processes even though the true saddles are not being used.

\item[3.] A final third route for defining nearness begins with the observation that even though the saddles for wave packets generally do not coalesce exactly where a caustic is encountered, there would typically be a near coalescence, i.e.~avoided crossing.   As parameters are varied, a saddle remains exposed until a caustic of some kind is encountered and it switches to hidden status beyond the caustic.  This would be expected to make the first route fail as well since a Newton-Raphson search beginning from a real trajectory would no longer have a unique saddle towards  which it would converge.  As one might have expected, the behavior of the real and imaginary parts of the saddle's action function changes character beyond a caustic.  Whereas, the real part of the action is varying rapidly along the exposed part of a foliation and the imaginary part relatively slowly, the opposite occurs for the hidden saddle, the imaginary part varies relatively rapidly, but not the real part.  The saddle's contribution switches from a rapidly varying phase to a rapidly diminishing magnitude as a function of the parameter.  In Figs.~\ref{ggwpd},\ref{ggwpdleft} it is clear where the transition between exposed and hidden occurs as the position is varied.

\end{itemize}

\subsection{Incorporating hidden saddles}
\label{sec:hs}

The semiclassical methods of~\cite{Vanvoorhis02,Pal16,Tomsovic18,Tomsovic18b} are designed to locate the physically relevant subset of exposed saddles.  The question which remains is how to identify physically relevant hidden saddles in cases where they must be found.  They are particularly relevant where exponentially decaying behavior is dominant, such as in tunneling problems, but even in the simple quartic oscillator example, there are locations where hidden saddles give contributions as large as those coming from the most relevant exposed saddles, as mentioned previously.  In such cases, they also need to be incorporated into the methods.  

The technique proposed here relies on the general feature that the saddles do not coalesce exactly in the neighborhood of caustics.  The idea is to begin with a complete set of exposed saddles, and then follow the complex saddles continuously as a function of the natural parameters.  Since the saddles do not collide precisely, they can be followed through and beyond the caustic regions without ambiguity.  As the parameters are changed slightly, the previous saddle is to be used as the input for a Newton-Raphson search instead of some real trajectory.  If as a parameter shifts, a particular saddle crosses the boundary between being exposed and hidden, the Newton-Raphson search using real trajectory input would fail whereas a search using the neighboring complex saddle as input has no relevance to the concept of nearness to real dynamics.  Thus, there is no convergence problem as one steps through the parameter variation.  This is illustrated in Fig.~\ref{contourfig} where the initial conditions for saddles associated with the foliations over a range of final position are shown.  For each near approach of two exposed saddles, which is near a turning point caustic in this case, the exposed saddles become hidden and one of them crosses a Stokes line.  The curvature of the avoided crossing introduces no problems to a Newton-Raphson search using the neighboring saddle as a starting point.

For the quartic oscillator this method identifies the most physically relevant hidden saddles, they all remain within the classical zone, and they even maintain a labelling with respect to the real classical foliations.  The homogeneity of the quartic oscillator Hamiltonian makes it a particularly simple case.  The problem of bald spots requiring complex time paths~\cite{Petersen14} does not arise for the quartic oscillator either.  Parenthetically, continuous parameter variation would also work for locating hidden saddles behind bald spots, but it adds an additional continuous variation variable and one has to be aware of when a branch cut is being crossed.  The idea of following exposed saddles through caustic regions is more general than just the turning point caustic example encountered here and extends to systems with more degrees of freedom, whatever their dynamics, integrable, chaotic or mixed~\cite{Pal16,Tomsovic18,Tomsovic18b}.   Nevertheless, even without considering the bald spot problem, the hidden saddles are certainly a more complicated story in general as even the double well creates a number of foreseeable complications.  For foliations whose trajectories pass near a caustic near the barrier multiple times in its history, it will be necessary to consider the crossing through the caustic (or barrier) at each instance, leading to a much larger multiplicity of hidden saddles.

\section{Conclusions}
\label{conclusions}

The semiclassical approximation can be extremely important in many physical contexts.  Of particular interest in this paper are problems in which part of the analysis concerns the propagation of Gaussian wave packets or coherent states in bosonic many-body systems.  For nonlinear dynamical systems, the saddle point approximation leads to a two-point boundary value problem with an infinity of solutions, almost all of which are physically irrelevant.  The practical problems with implementing the theory only mount further as a system's number of degrees of freedom increases or the dynamics become chaotic.  Two techniques, i.e.~off-center real trajectory methods~\cite{Tomsovic93,Barnes94} and using those resultant trajectories to identify physically relevant saddles~\cite{Vanvoorhis02,Pal16,Tomsovic18,Tomsovic18b}, are placed on a firmer foundation by showing how they are related to dynamical structure that partitions the Lagrangian manifolds underlying wave packet and coherent state propagation.

For a simple one degree of freedom dynamical system, the pure quartic oscillator, the wave packet / coherent state related Lagrangian manifolds can be partitioned into an infinite number of subregions by the lines of initial conditions of singular trajectories.  One region in particular, here dubbed the classical zone, contains all the exposed and hidden saddles discussed here.  The exposed saddles can be placed in a one-to-one correspondence with all the real classical transport pathways, and are straightforwardly located with a Newton-Raphson search technique fed by a single reference trajectory from each transport pathway.  A simple criterion based on the real reference trajectory's initial conditions and final phase point can be used to determine which saddles contribute sufficiently for them to be considered physically relevant.  It only involves the differences between the initial phase space centroid and the initial conditions, and the final phase point and final centroid.

The hidden saddles, where relevant, cannot be identified this way.  However, one can begin with the maximal set of exposed saddles (for the quartic oscillator this means setting the final position to zero) and vary the relevant parameters smoothly, here position.  Following each exposed saddle continuously with final position leads eventually to a regime where that saddle cannot be found with a Newton-Raphson search using a real trajectory as initial input.  Nevertheless, one can follow each saddle as it moves from an exposed region into a hidden one.  Some of the hidden saddles found this way do cross Stokes lines and need to be thrown away, but that can be evaluated by calculating their complex action functions.  For this simple dynamical example, all the physically relevant hidden saddles necessary to calculate the propagating wave packet / coherent state out in the exponentially decreasing tail regions are from the classical zone and could be found through real parameter variation of final position.  This method would work in an identical way if the quantity of interest were the overlap of the propagated wave function with a final coherent state/ wave packet.  The parameter(s) varied would be the centroid of the final wave packet.

The dynamics of the pure quartic oscillator are extremely simple, not only because it is a one-degree-of-freedom system, but also because the potential is homogeneous.  This gives rise to simple scaling relations between trajectories on any pair of energy surfaces.  More general dynamical systems might be expected to give rise to further challenges not encountered in this simple example.  Indeed, one issue that has been identified is the bald spot problem~\cite{Huber88, Petersen14}.  This arises when a system possesses movable branch points as a function of initial conditions, which in turn may block the Newton-Raphson search scheme for saddles beyond some point as dynamical quantities are varied (such as the position variable in the propagated wave function).  In these circumstances, complex time contours must be introduced in order to circumvent the branch points and locate the saddles ``behind'' the bald spots~\cite{Petersen14}.  Thus, depending on the system, complex time may be unavoidable.

Further complications may arise with the introduction of multiple degrees of freedom, where the dynamics may include KAM tori, chaotic regions, Arnol'd diffusion, etc...   The off-center trajectory and exposed saddle methods have been shown to work in higher dimensional systems.  Nevertheless, it is not known whether the partitioning of the Lagrangian manifold found here has a straightforward multi-degree of freedom generalization nor is it known whether the appearance of chaotic dynamics fundamentally alters the picture of a partitioned Lagrangian manifold.  One natural extension of the current work would be to investigate the possibility of such higher dimensional partitionings.

\bibliography{quantumchaos,manybody,general_ref,molecular}

\begin{thebibliography}{34}%
\makeatletter
\providecommand \@ifxundefined [1]{%
 \@ifx{#1\undefined}
}%
\providecommand \@ifnum [1]{%
 \ifnum #1\expandafter \@firstoftwo
 \else \expandafter \@secondoftwo
 \fi
}%
\providecommand \@ifx [1]{%
 \ifx #1\expandafter \@firstoftwo
 \else \expandafter \@secondoftwo
 \fi
}%
\providecommand \natexlab [1]{#1}%
\providecommand \enquote  [1]{``#1''}%
\providecommand \bibnamefont  [1]{#1}%
\providecommand \bibfnamefont [1]{#1}%
\providecommand \citenamefont [1]{#1}%
\providecommand \href@noop [0]{\@secondoftwo}%
\providecommand \href [0]{\begingroup \@sanitize@url \@href}%
\providecommand \@href[1]{\@@startlink{#1}\@@href}%
\providecommand \@@href[1]{\endgroup#1\@@endlink}%
\providecommand \@sanitize@url [0]{\catcode `\\12\catcode `\$12\catcode
  `\&12\catcode `\#12\catcode `\^12\catcode `\_12\catcode `\%12\relax}%
\providecommand \@@startlink[1]{}%
\providecommand \@@endlink[0]{}%
\providecommand \url  [0]{\begingroup\@sanitize@url \@url }%
\providecommand \@url [1]{\endgroup\@href {#1}{\urlprefix }}%
\providecommand \urlprefix  [0]{URL }%
\providecommand \Eprint [0]{\href }%
\providecommand \doibase [0]{https://doi.org/}%
\providecommand \selectlanguage [0]{\@gobble}%
\providecommand \bibinfo  [0]{\@secondoftwo}%
\providecommand \bibfield  [0]{\@secondoftwo}%
\providecommand \translation [1]{[#1]}%
\providecommand \BibitemOpen [0]{}%
\providecommand \bibitemStop [0]{}%
\providecommand \bibitemNoStop [0]{.\EOS\space}%
\providecommand \EOS [0]{\spacefactor3000\relax}%
\providecommand \BibitemShut  [1]{\csname bibitem#1\endcsname}%
\let\auto@bib@innerbib\@empty
\bibitem [{\citenamefont {Glauber}(1963)}]{Glauber63}%
  \BibitemOpen
  \bibfield  {author} {\bibinfo {author} {\bibfnamefont {R.~J.}\ \bibnamefont
  {Glauber}},\ }\bibfield  {title} {\bibinfo {title} {Coherent and incoherent
  states of the radiation field},\ }\href@noop {} {\bibfield  {journal}
  {\bibinfo  {journal} {Phys.~Rev.}\ }\textbf {\bibinfo {volume} {131}},\
  \bibinfo {pages} {2766} (\bibinfo {year} {1963})}\BibitemShut {NoStop}%
\bibitem [{\citenamefont {Scully}\ and\ \citenamefont
  {Zubairy}(1997)}]{Scully97}%
  \BibitemOpen
  \bibfield  {author} {\bibinfo {author} {\bibfnamefont {M.~O.}\ \bibnamefont
  {Scully}}\ and\ \bibinfo {author} {\bibfnamefont {M.~S.}\ \bibnamefont
  {Zubairy}},\ }\href@noop {} {\emph {\bibinfo {title} {Quantum Optics}}}\
  (\bibinfo  {publisher} {Cambridge University Press},\ \bibinfo {address}
  {Cambridge, UK},\ \bibinfo {year} {1997})\BibitemShut {NoStop}%
\bibitem [{\citenamefont {Greiner}\ \emph {et~al.}(2002)\citenamefont
  {Greiner}, \citenamefont {Mandel}, \citenamefont {H\"ansch},\ and\
  \citenamefont {Bloch}}]{Greiner02b}%
  \BibitemOpen
  \bibfield  {author} {\bibinfo {author} {\bibfnamefont {M.}~\bibnamefont
  {Greiner}}, \bibinfo {author} {\bibfnamefont {O.}~\bibnamefont {Mandel}},
  \bibinfo {author} {\bibfnamefont {T.~W.}\ \bibnamefont {H\"ansch}},\ and\
  \bibinfo {author} {\bibfnamefont {I.}~\bibnamefont {Bloch}},\ }\bibfield
  {title} {\bibinfo {title} {Collapse and revival of the matter wave field of a
  bose-einstein condensate},\ }\href@noop {} {\bibfield  {journal} {\bibinfo
  {journal} {Nature}\ }\textbf {\bibinfo {volume} {419}},\ \bibinfo {pages}
  {51} (\bibinfo {year} {2002})}\BibitemShut {NoStop}%
\bibitem [{\citenamefont {Polkovnikov}\ \emph {et~al.}(2011)\citenamefont
  {Polkovnikov}, \citenamefont {Sengupta}, \citenamefont {Silva},\ and\
  \citenamefont {Vengalattore}}]{Polkovnikov11}%
  \BibitemOpen
  \bibfield  {author} {\bibinfo {author} {\bibfnamefont {A.}~\bibnamefont
  {Polkovnikov}}, \bibinfo {author} {\bibfnamefont {K.}~\bibnamefont
  {Sengupta}}, \bibinfo {author} {\bibfnamefont {A.}~\bibnamefont {Silva}},\
  and\ \bibinfo {author} {\bibfnamefont {M.}~\bibnamefont {Vengalattore}},\
  }\bibfield  {title} {\bibinfo {title} {Colloquium: Nonequilibrium dynamics of
  closed interacting quantum systems},\ }\href@noop {} {\bibfield  {journal}
  {\bibinfo  {journal} {Rev.~Mod.~Phys.}\ }\textbf {\bibinfo {volume} {83}},\
  \bibinfo {pages} {863} (\bibinfo {year} {2011})}\BibitemShut {NoStop}%
\bibitem [{\citenamefont {Heller}(1981)}]{Heller81b}%
  \BibitemOpen
  \bibfield  {author} {\bibinfo {author} {\bibfnamefont {E.~J.}\ \bibnamefont
  {Heller}},\ }\bibfield  {title} {\bibinfo {title} {The semiclassical way to
  molecular spectroscopy},\ }\href@noop {} {\bibfield  {journal} {\bibinfo
  {journal} {Acc.~Chem.~Res.}\ }\textbf {\bibinfo {volume} {14}},\ \bibinfo
  {pages} {368} (\bibinfo {year} {1981})}\BibitemShut {NoStop}%
\bibitem [{\citenamefont {Gruebele}\ and\ \citenamefont
  {Zewail}(1992)}]{Gruebele92}%
  \BibitemOpen
  \bibfield  {author} {\bibinfo {author} {\bibfnamefont {M.}~\bibnamefont
  {Gruebele}}\ and\ \bibinfo {author} {\bibfnamefont {A.~H.}\ \bibnamefont
  {Zewail}},\ }\bibfield  {title} {\bibinfo {title} {Femtosecond wave packet
  spectroscopy: Coherences, the potential, and structural determination},\
  }\href@noop {} {\bibfield  {journal} {\bibinfo  {journal} {J.~Chem.~Phys.}\
  }\textbf {\bibinfo {volume} {98}},\ \bibinfo {pages} {883} (\bibinfo {year}
  {1992})}\BibitemShut {NoStop}%
\bibitem [{\citenamefont {Zewail}(2000)}]{Zewail00}%
  \BibitemOpen
  \bibfield  {author} {\bibinfo {author} {\bibfnamefont {A.~H.}\ \bibnamefont
  {Zewail}},\ }\bibfield  {title} {\bibinfo {title} {Femtochemistry:
  Atomic-scale dynamics of the chemical bond},\ }\href@noop {} {\bibfield
  {journal} {\bibinfo  {journal} {J.~Phys.~Chem.~A}\ }\textbf {\bibinfo
  {volume} {104}},\ \bibinfo {pages} {5660} (\bibinfo {year}
  {2000})}\BibitemShut {NoStop}%
\bibitem [{\citenamefont {Agostini}\ and\ \citenamefont
  {DiMauro}(2004)}]{Agostini04}%
  \BibitemOpen
  \bibfield  {author} {\bibinfo {author} {\bibfnamefont {P.}~\bibnamefont
  {Agostini}}\ and\ \bibinfo {author} {\bibfnamefont {L.~F.}\ \bibnamefont
  {DiMauro}},\ }\bibfield  {title} {\bibinfo {title} {The physics of attosecond
  light pulses},\ }\href@noop {} {\bibfield  {journal} {\bibinfo  {journal}
  {Rep.~Prog.~Phys.}\ }\textbf {\bibinfo {volume} {67}},\ \bibinfo {pages}
  {813} (\bibinfo {year} {2004})}\BibitemShut {NoStop}%
\bibitem [{\citenamefont {Klauder}\ and\ \citenamefont
  {Skagerstam}(1985)}]{Klauder85}%
  \BibitemOpen
  \bibfield  {author} {\bibinfo {author} {\bibfnamefont {J.~R.}\ \bibnamefont
  {Klauder}}\ and\ \bibinfo {author} {\bibfnamefont {B.-S.}\ \bibnamefont
  {Skagerstam}},\ }\href@noop {} {\emph {\bibinfo {title} {Coherent States:
  Applications in Physics and Mathematical Physics}}}\ (\bibinfo  {publisher}
  {World Scientific},\ \bibinfo {address} {Singapore},\ \bibinfo {year}
  {1985})\BibitemShut {NoStop}%
\bibitem [{\citenamefont {Huber}\ \emph {et~al.}(1988)\citenamefont {Huber},
  \citenamefont {Heller},\ and\ \citenamefont {Littlejohn}}]{Huber88}%
  \BibitemOpen
  \bibfield  {author} {\bibinfo {author} {\bibfnamefont {D.}~\bibnamefont
  {Huber}}, \bibinfo {author} {\bibfnamefont {E.~J.}\ \bibnamefont {Heller}},\
  and\ \bibinfo {author} {\bibfnamefont {R.~G.}\ \bibnamefont {Littlejohn}},\
  }\bibfield  {title} {\bibinfo {title} {Generalized gaussian wave packet
  dynamics, schr\"odinger equation, and stationary phase approximation},\
  }\href@noop {} {\bibfield  {journal} {\bibinfo  {journal} {J.~Chem.~Phys.}\
  }\textbf {\bibinfo {volume} {89}},\ \bibinfo {pages} {2003} (\bibinfo {year}
  {1988})}\BibitemShut {NoStop}%
\bibitem [{\citenamefont {Baranger}\ \emph {et~al.}(2001)\citenamefont
  {Baranger}, \citenamefont {de~Aguiar}, \citenamefont {Keck}, \citenamefont
  {Korsch},\ and\ \citenamefont {Schellhaass}}]{Baranger01}%
  \BibitemOpen
  \bibfield  {author} {\bibinfo {author} {\bibfnamefont {M.}~\bibnamefont
  {Baranger}}, \bibinfo {author} {\bibfnamefont {M.~A.~M.}\ \bibnamefont
  {de~Aguiar}}, \bibinfo {author} {\bibfnamefont {F.}~\bibnamefont {Keck}},
  \bibinfo {author} {\bibfnamefont {H.~J.}\ \bibnamefont {Korsch}},\ and\
  \bibinfo {author} {\bibfnamefont {B.}~\bibnamefont {Schellhaass}},\
  }\bibfield  {title} {\bibinfo {title} {Semiclassical approximations in phase
  space with coherent states},\ }\href@noop {} {\bibfield  {journal} {\bibinfo
  {journal} {J.~Phys.~A:~Math.~Gen.}\ }\textbf {\bibinfo {volume} {34}},\
  \bibinfo {pages} {7227} (\bibinfo {year} {2001})}\BibitemShut {NoStop}%
\bibitem [{\citenamefont {Berry}\ and\ \citenamefont
  {Balazs}(1979)}]{Berry79b}%
  \BibitemOpen
  \bibfield  {author} {\bibinfo {author} {\bibfnamefont {M.~V.}\ \bibnamefont
  {Berry}}\ and\ \bibinfo {author} {\bibfnamefont {N.~L.}\ \bibnamefont
  {Balazs}},\ }\bibfield  {title} {\bibinfo {title} {Evolution of semiclassical
  quantum states in phase space},\ }\href@noop {} {\bibfield  {journal}
  {\bibinfo  {journal} {J.~Phys.~A}\ }\textbf {\bibinfo {volume} {12}},\
  \bibinfo {pages} {625} (\bibinfo {year} {1979})}\BibitemShut {NoStop}%
\bibitem [{\citenamefont {Heller}(1975)}]{Heller75}%
  \BibitemOpen
  \bibfield  {author} {\bibinfo {author} {\bibfnamefont {E.~J.}\ \bibnamefont
  {Heller}},\ }\bibfield  {title} {\bibinfo {title} {Time-dependent approach to
  semiclassical dynamics},\ }\href@noop {} {\bibfield  {journal} {\bibinfo
  {journal} {J.~Chem.~Phys.}\ }\textbf {\bibinfo {volume} {62}},\ \bibinfo
  {pages} {1544} (\bibinfo {year} {1975})}\BibitemShut {NoStop}%
\bibitem [{\citenamefont {Heller}\ and\ \citenamefont
  {Tomsovic}(1993)}]{Heller93}%
  \BibitemOpen
  \bibfield  {author} {\bibinfo {author} {\bibfnamefont {E.~J.}\ \bibnamefont
  {Heller}}\ and\ \bibinfo {author} {\bibfnamefont {S.}~\bibnamefont
  {Tomsovic}},\ }\bibfield  {title} {\bibinfo {title} {Post-modern quantum
  mechanics},\ }\href@noop {} {\bibfield  {journal} {\bibinfo  {journal}
  {Physics Today}\ }\textbf {\bibinfo {volume} {46}},\ \bibinfo {pages} {38}
  (\bibinfo {year} {1993})},\ \bibinfo {note} {reprinted in "Parity", Vol. 12
  (8), (1993); Figures 1 \& 6 reproduced in "Semiclassical Physics", Brack and
  Bhaduri, Addison-Wesley, (1997, USA)}\BibitemShut {NoStop}%
\bibitem [{\citenamefont {Tomsovic}\ \emph {et~al.}(2018)\citenamefont
  {Tomsovic}, \citenamefont {Schlagheck}, \citenamefont {Ullmo}, \citenamefont
  {Urbina},\ and\ \citenamefont {Richter}}]{Tomsovic18}%
  \BibitemOpen
  \bibfield  {author} {\bibinfo {author} {\bibfnamefont {S.}~\bibnamefont
  {Tomsovic}}, \bibinfo {author} {\bibfnamefont {P.}~\bibnamefont
  {Schlagheck}}, \bibinfo {author} {\bibfnamefont {D.}~\bibnamefont {Ullmo}},
  \bibinfo {author} {\bibfnamefont {J.-D.}\ \bibnamefont {Urbina}},\ and\
  \bibinfo {author} {\bibfnamefont {K.}~\bibnamefont {Richter}},\ }\bibfield
  {title} {\bibinfo {title} {Post-ehrenfest many-body quantum interferences in
  ultracold atoms far-out-of-equilibrium},\ }\href@noop {} {\bibfield
  {journal} {\bibinfo  {journal} {Phys.~Rev.~A}\ }\textbf {\bibinfo {volume}
  {97}},\ \bibinfo {pages} {061606(R)} (\bibinfo {year} {2018})},\ \bibinfo
  {note} {arXiv:1711.04693v2 [quant-ph]}\BibitemShut {NoStop}%
\bibitem [{\citenamefont {Ehrenfest}(1927)}]{Ehrenfest27}%
  \BibitemOpen
  \bibfield  {author} {\bibinfo {author} {\bibfnamefont {P.}~\bibnamefont
  {Ehrenfest}},\ }\bibfield  {title} {\bibinfo {title} {Bemerkung \"uber die
  angen\"aherte g\"ultigkeit der klassischen mechanik innerhalb der
  quantenmechanik},\ }\href@noop {} {\bibfield  {journal} {\bibinfo  {journal}
  {Zeit.~Phys.}\ }\textbf {\bibinfo {volume} {45}},\ \bibinfo {pages} {455}
  (\bibinfo {year} {1927})}\BibitemShut {NoStop}%
\bibitem [{\citenamefont {Wang}\ and\ \citenamefont
  {Tomsovic}(2021)}]{Wang21b}%
  \BibitemOpen
  \bibfield  {author} {\bibinfo {author} {\bibfnamefont {H.}~\bibnamefont
  {Wang}}\ and\ \bibinfo {author} {\bibfnamefont {S.}~\bibnamefont
  {Tomsovic}},\ }\bibfield  {title} {\bibinfo {title} {A corrected maslov index
  for complex saddle trajectories},\ }\href@noop {} {\bibfield  {journal}
  {\bibinfo  {journal} {arXiv:2108.04301 [quant-ph]}\ } (\bibinfo {year}
  {2021})}\BibitemShut {NoStop}%
\bibitem [{\citenamefont {Van~Voorhis}\ and\ \citenamefont
  {Heller}(2002)}]{Vanvoorhis02}%
  \BibitemOpen
  \bibfield  {author} {\bibinfo {author} {\bibfnamefont {T.}~\bibnamefont
  {Van~Voorhis}}\ and\ \bibinfo {author} {\bibfnamefont {E.~J.}\ \bibnamefont
  {Heller}},\ }\bibfield  {title} {\bibinfo {title} {Nearly real trajectories
  in complex semiclassical dynamics},\ }\href@noop {} {\bibfield  {journal}
  {\bibinfo  {journal} {Phys.~Rev.~A}\ }\textbf {\bibinfo {volume} {66}},\
  \bibinfo {pages} {050501(R)} (\bibinfo {year} {2002})}\BibitemShut {NoStop}%
\bibitem [{\citenamefont {Van~Voorhis}\ and\ \citenamefont
  {Heller}(2003)}]{Vanvoorhis03}%
  \BibitemOpen
  \bibfield  {author} {\bibinfo {author} {\bibfnamefont {T.}~\bibnamefont
  {Van~Voorhis}}\ and\ \bibinfo {author} {\bibfnamefont {E.~J.}\ \bibnamefont
  {Heller}},\ }\bibfield  {title} {\bibinfo {title} {Similarity transformed
  semiclassical dynamics},\ }\href@noop {} {\bibfield  {journal} {\bibinfo
  {journal} {J.~Chem.~Phys.}\ }\textbf {\bibinfo {volume} {119}},\ \bibinfo
  {pages} {12153} (\bibinfo {year} {2003})}\BibitemShut {NoStop}%
\bibitem [{\citenamefont {Pal}\ \emph {et~al.}(2016)\citenamefont {Pal},
  \citenamefont {Vyas},\ and\ \citenamefont {Tomsovic}}]{Pal16}%
  \BibitemOpen
  \bibfield  {author} {\bibinfo {author} {\bibfnamefont {H.}~\bibnamefont
  {Pal}}, \bibinfo {author} {\bibfnamefont {M.}~\bibnamefont {Vyas}},\ and\
  \bibinfo {author} {\bibfnamefont {S.}~\bibnamefont {Tomsovic}},\ }\bibfield
  {title} {\bibinfo {title} {Generalized gaussian wave packet dynamics:
  Integrable and chaotic systems},\ }\href@noop {} {\bibfield  {journal}
  {\bibinfo  {journal} {Phys.~Rev.~E}\ }\textbf {\bibinfo {volume} {93}},\
  \bibinfo {pages} {012213} (\bibinfo {year} {2016})},\ \bibinfo {note}
  {arXiv:1510.08051 [quant-ph]}\BibitemShut {NoStop}%
\bibitem [{\citenamefont {Tomsovic}(2018)}]{Tomsovic18b}%
  \BibitemOpen
  \bibfield  {author} {\bibinfo {author} {\bibfnamefont {S.}~\bibnamefont
  {Tomsovic}},\ }\bibfield  {title} {\bibinfo {title} {Complex saddle
  trajectories for multidimensional quantum wave packet/coherent state
  propagation: application to a many-body system},\ }\href@noop {} {\bibfield
  {journal} {\bibinfo  {journal} {Phys.~Rev.~E}\ }\textbf {\bibinfo {volume}
  {98}},\ \bibinfo {pages} {023301} (\bibinfo {year} {2018})},\ \bibinfo {note}
  {arXiv:1804.10511 [cond-mat.stat-mech]}\BibitemShut {NoStop}%
\bibitem [{\citenamefont {O'Connor}\ \emph {et~al.}(1992)\citenamefont
  {O'Connor}, \citenamefont {Tomsovic},\ and\ \citenamefont
  {Heller}}]{Oconnor92}%
  \BibitemOpen
  \bibfield  {author} {\bibinfo {author} {\bibfnamefont {P.~W.}\ \bibnamefont
  {O'Connor}}, \bibinfo {author} {\bibfnamefont {S.}~\bibnamefont {Tomsovic}},\
  and\ \bibinfo {author} {\bibfnamefont {E.~J.}\ \bibnamefont {Heller}},\
  }\bibfield  {title} {\bibinfo {title} {Semiclassical dynamics in the strongly
  chaotic regime - breaking the log-time barrier},\ }\href@noop {} {\bibfield
  {journal} {\bibinfo  {journal} {Physica D}\ }\textbf {\bibinfo {volume}
  {55}},\ \bibinfo {pages} {340} (\bibinfo {year} {1992})}\BibitemShut
  {NoStop}%
\bibitem [{\citenamefont {Tomsovic}\ and\ \citenamefont
  {Heller}(1993{\natexlab{a}})}]{Tomsovic93}%
  \BibitemOpen
  \bibfield  {author} {\bibinfo {author} {\bibfnamefont {S.}~\bibnamefont
  {Tomsovic}}\ and\ \bibinfo {author} {\bibfnamefont {E.~J.}\ \bibnamefont
  {Heller}},\ }\bibfield  {title} {\bibinfo {title} {The long-time
  semiclassical dynamics of chaos: the stadium billiard},\ }\href@noop {}
  {\bibfield  {journal} {\bibinfo  {journal} {Phys.~Rev.~E}\ }\textbf {\bibinfo
  {volume} {47}},\ \bibinfo {pages} {282} (\bibinfo {year}
  {1993}{\natexlab{a}})}\BibitemShut {NoStop}%
\bibitem [{\citenamefont {Tomsovic}\ and\ \citenamefont
  {Heller}(1993{\natexlab{b}})}]{Tomsovic93b}%
  \BibitemOpen
  \bibfield  {author} {\bibinfo {author} {\bibfnamefont {S.}~\bibnamefont
  {Tomsovic}}\ and\ \bibinfo {author} {\bibfnamefont {E.~J.}\ \bibnamefont
  {Heller}},\ }\bibfield  {title} {\bibinfo {title} {The semiclassical
  construction of chaotic eigenstates},\ }\href@noop {} {\bibfield  {journal}
  {\bibinfo  {journal} {Phys.~Rev.~Lett.}\ }\textbf {\bibinfo {volume} {70}},\
  \bibinfo {pages} {1405} (\bibinfo {year} {1993}{\natexlab{b}})}\BibitemShut
  {NoStop}%
\bibitem [{\citenamefont {Barnes}\ \emph {et~al.}(1994)\citenamefont {Barnes},
  \citenamefont {Nauenberg}, \citenamefont {Nockleby},\ and\ \citenamefont
  {Tomsovic}}]{Barnes94}%
  \BibitemOpen
  \bibfield  {author} {\bibinfo {author} {\bibfnamefont {I.~M.~S.}\
  \bibnamefont {Barnes}}, \bibinfo {author} {\bibfnamefont {M.}~\bibnamefont
  {Nauenberg}}, \bibinfo {author} {\bibfnamefont {M.}~\bibnamefont
  {Nockleby}},\ and\ \bibinfo {author} {\bibfnamefont {S.}~\bibnamefont
  {Tomsovic}},\ }\bibfield  {title} {\bibinfo {title} {Classical orbits and
  semiclassical wave packet propagation in the coulomb potential},\ }\href@noop
  {} {\bibfield  {journal} {\bibinfo  {journal} {J.~Phys.~A: Math.~Gen.}\
  }\textbf {\bibinfo {volume} {27}},\ \bibinfo {pages} {3299} (\bibinfo {year}
  {1994})}\BibitemShut {NoStop}%
\bibitem [{\citenamefont {Maslov}\ and\ \citenamefont
  {Fedoriuk}(1981)}]{Maslov81}%
  \BibitemOpen
  \bibfield  {author} {\bibinfo {author} {\bibfnamefont {V.~P.}\ \bibnamefont
  {Maslov}}\ and\ \bibinfo {author} {\bibfnamefont {M.~V.}\ \bibnamefont
  {Fedoriuk}},\ }\href@noop {} {\emph {\bibinfo {title} {Semiclassical
  approximation in quantum mechanics}}}\ (\bibinfo  {publisher} {Reidel
  Publishing Company},\ \bibinfo {address} {Dordrecht},\ \bibinfo {year}
  {1981})\BibitemShut {NoStop}%
\bibitem [{\citenamefont {Creagh}(1998)}]{Creagh98}%
  \BibitemOpen
  \bibfield  {author} {\bibinfo {author} {\bibfnamefont {S.~C.}\ \bibnamefont
  {Creagh}},\ }\bibfield  {title} {\bibinfo {title} {Tunneling in two
  dimensions},\ }in\ \href@noop {} {\emph {\bibinfo {booktitle} {Tunneling in
  complex systems, Proceedings from the Institute for Nuclear Theory: Volume
  5}}},\ \bibinfo {editor} {edited by\ \bibinfo {editor} {\bibfnamefont
  {S.}~\bibnamefont {Tomsovic}}}\ (\bibinfo  {publisher} {World Scientific},\
  \bibinfo {address} {Singapore},\ \bibinfo {year} {1998})\ pp.\ \bibinfo
  {pages} {35--100}\BibitemShut {NoStop}%
\bibitem [{\citenamefont {Petersen}\ and\ \citenamefont
  {Kay}(2014)}]{Petersen14}%
  \BibitemOpen
  \bibfield  {author} {\bibinfo {author} {\bibfnamefont {J.}~\bibnamefont
  {Petersen}}\ and\ \bibinfo {author} {\bibfnamefont {K.~G.}\ \bibnamefont
  {Kay}},\ }\bibfield  {title} {\bibinfo {title} {Complex time paths for
  semiclassical wave packet propagation with complex trajectories},\
  }\href@noop {} {\bibfield  {journal} {\bibinfo  {journal} {J.~Chem.~Phys.}\
  }\textbf {\bibinfo {volume} {141}},\ \bibinfo {pages} {054114} (\bibinfo
  {year} {2014})}\BibitemShut {NoStop}%
\bibitem [{\citenamefont {Tomsovic}\ and\ \citenamefont
  {Heller}(1991)}]{Tomsovic91b}%
  \BibitemOpen
  \bibfield  {author} {\bibinfo {author} {\bibfnamefont {S.}~\bibnamefont
  {Tomsovic}}\ and\ \bibinfo {author} {\bibfnamefont {E.~J.}\ \bibnamefont
  {Heller}},\ }\bibfield  {title} {\bibinfo {title} {Semiclassical dynamics of
  chaotic motion: Unexpected long time accuracy},\ }\href@noop {} {\bibfield
  {journal} {\bibinfo  {journal} {Phys.~Rev.~Lett.}\ }\textbf {\bibinfo
  {volume} {67}},\ \bibinfo {pages} {664} (\bibinfo {year} {1991})}\BibitemShut
  {NoStop}%
\bibitem [{\citenamefont {Barnes}\ \emph {et~al.}(1993)\citenamefont {Barnes},
  \citenamefont {Nauenberg}, \citenamefont {Nockleby},\ and\ \citenamefont
  {Tomsovic}}]{Barnes93}%
  \BibitemOpen
  \bibfield  {author} {\bibinfo {author} {\bibfnamefont {I.~M.~S.}\
  \bibnamefont {Barnes}}, \bibinfo {author} {\bibfnamefont {M.}~\bibnamefont
  {Nauenberg}}, \bibinfo {author} {\bibfnamefont {M.}~\bibnamefont
  {Nockleby}},\ and\ \bibinfo {author} {\bibfnamefont {S.}~\bibnamefont
  {Tomsovic}},\ }\bibfield  {title} {\bibinfo {title} {Semiclassical theory of
  quantum propagation: The coulomb potential},\ }\href@noop {} {\bibfield
  {journal} {\bibinfo  {journal} {Phys.~Rev.~Lett.}\ }\textbf {\bibinfo
  {volume} {71}},\ \bibinfo {pages} {1961} (\bibinfo {year}
  {1993})}\BibitemShut {NoStop}%
\bibitem [{\citenamefont {Sep\'ulveda}\ \emph {et~al.}(1992)\citenamefont
  {Sep\'ulveda}, \citenamefont {Tomsovic},\ and\ \citenamefont
  {Heller}}]{Sepulveda92}%
  \BibitemOpen
  \bibfield  {author} {\bibinfo {author} {\bibfnamefont {M.-A.}\ \bibnamefont
  {Sep\'ulveda}}, \bibinfo {author} {\bibfnamefont {S.}~\bibnamefont
  {Tomsovic}},\ and\ \bibinfo {author} {\bibfnamefont {E.~J.}\ \bibnamefont
  {Heller}},\ }\bibfield  {title} {\bibinfo {title} {Semiclassical propagation:
  how long can it last},\ }\href@noop {} {\bibfield  {journal} {\bibinfo
  {journal} {Phys.~Rev.~Lett.}\ }\textbf {\bibinfo {volume} {69}},\ \bibinfo
  {pages} {402} (\bibinfo {year} {1992})}\BibitemShut {NoStop}%
\bibitem [{\citenamefont {Ince}(1956)}]{Ince56}%
  \BibitemOpen
  \bibfield  {author} {\bibinfo {author} {\bibfnamefont {E.~L.}\ \bibnamefont
  {Ince}},\ }\href@noop {} {\emph {\bibinfo {title} {Ordinary Differential
  Equations}}}\ (\bibinfo  {publisher} {Dover},\ \bibinfo {address} {Mineola,
  New York},\ \bibinfo {year} {1956})\BibitemShut {NoStop}%
\bibitem [{\citenamefont {Polyanin}\ and\ \citenamefont
  {Zaitsev}(2003)}]{Polyanin03}%
  \BibitemOpen
  \bibfield  {author} {\bibinfo {author} {\bibfnamefont {A.~D.}\ \bibnamefont
  {Polyanin}}\ and\ \bibinfo {author} {\bibfnamefont {V.~F.}\ \bibnamefont
  {Zaitsev}},\ }\href@noop {} {\emph {\bibinfo {title} {Handbook of exact
  solutions for ordinary differential equations}}}\ (\bibinfo  {publisher}
  {Chapman and Hall/CRC},\ \bibinfo {address} {Boca Raton},\ \bibinfo {year}
  {2003})\BibitemShut {NoStop}%
\bibitem [{\citenamefont {Berry}(1989)}]{Berry89}%
  \BibitemOpen
  \bibfield  {author} {\bibinfo {author} {\bibfnamefont {M.~V.}\ \bibnamefont
  {Berry}},\ }\bibfield  {title} {\bibinfo {title} {Uniform asymptotic
  smoothing of stokes' discontinuities},\ }\href@noop {} {\bibfield  {journal}
  {\bibinfo  {journal} {Proc.~R.~Soc.~A}\ }\textbf {\bibinfo {volume} {422}},\
  \bibinfo {pages} {7} (\bibinfo {year} {1989})}\BibitemShut {NoStop}%
\end{thebibliography}%

\end{document}